\documentclass[reqno,english,aps,prb,superscriptaddress,bibliographystyle=apsrev4-]{revtex4-2}
\usepackage{amsmath}
\usepackage{amsthm}
\usepackage{amssymb}
\usepackage{accents}
\usepackage{fontspec}

\usepackage{unicode-math}

\usepackage{xcolor}

\usepackage{mathtools}
\usepackage{float}
\usepackage{graphicx}
\usepackage{geometry}
\usepackage[title]{appendix}

\usepackage{microtype}

\newcommand{\umacron}{\Umathchar 0 \symoperators "00AF}
\newcommand{\ubar}[1]{\underaccent{\textstyle{\umacron}}{#1}}
\newcommand{\longleft}{\hspace{2em}&\hspace{-2em}}

\usepackage[nameinlink,capitalize]{cleveref}
\crefname{figure}{Figure}{Figures}
\Crefname{figure}{Figure}{Figures}

\AtBeginDocument{
\global\long\def\bwhat#1{\widehat{#1}}%
\global\long\def\vc#1{\symbfit{#1}}%
\global\long\def\Sqrt#1{\sqrt{#1}}%
\global\long\def\pp{\uppi}%
\global\long\def\ii{\mathrm{i}}%
\global\long\def\kd#1{\mathsf{\delta}_{#1}}%
\global\long\def\lhs#1{\longleft#1}%
\global\long\def\bsum{\sum}%
\global\long\def\uscr#1#2{\mathscr{e}_{\mathcal{#1}#2}}%
\global\long\def\uvcr#1{\hat{\mathbfscr{e}}_{\mathcal{#1}}}%
\global\long\def\tr{\operatorname{tr}}%
\global\long\def\*{\!\!}%
\global\long\def\d#1{\operatorname{\symrm{d}\symit{#1}}}
\global\long\def\Int#1#2{\int\displaylimits^{#2}_{#1}}%
\global\long\def\ocurl#1{\mathscr{#1}}%
\global\long\def\nwcal#1{\mathscr{#1}}%
\global\long\def\ub#1{\ubar{#1}}%
\global\long\def\uv#1{\hat{\mathbf{e}}_{#1}}%
\global\long\def\mts#1{\mathcal{#1}}%
\global\long\def\sfit#1{\mathsfit{#1}}%
}
\begin{document}
\title{Polarized light Raman scattering by an atom near\\ an ultrathin periodically aligned carbon nanotube film}
\author{SK Firoz Islam}
\affiliation{Department of Mathematics and Physics, North Carolina Central University,
	Durham, North Carolina 27707, USA}
\affiliation{Department of Physics, Jamia Millia Islamia (A Central University),
	New Delhi 110025, INDIA}
\author{Michael Dean Pugh}
\affiliation{Department of Mathematics and Physics, North Carolina Central University,
	Durham, North Carolina 27707, USA}
\author{Igor V. Bondarev}
\affiliation{Department of Mathematics and Physics, North Carolina Central University,
	Durham, North Carolina 27707, USA}

\begin{abstract}
	We present a systematic theoretical study of the Raman scattering effect for a two-level atomic system in near proximity of an ultrathin dielectric film with an embedded parallel array of periodically aligned single-wall semiconducting carbon nanotubes. More generally, our model provides a unified description of the quantum near-field medium-assisted enhancement effects for in-plane anisotropic metasurfaces, of which ultrathin periodically aligned carbon nanotube films are the representative example. Particular attention is given to incoming photon parameters of the external light radiation such as polarization and incidence plane orientation relative to the main anisotropy axis (nanotube alignment axis). By explicitly deriving the Raman scattering cross-section, we establish that for the two-level atomic system in the near-field zone of the carbon nanotube metasurface the effect can be enhanced by a factor of up to $10^{4}$, not only for $p$-polarized but for $s$-polarized light as well.
\end{abstract}

\maketitle

\section{Introduction}

Surface-enhanced Raman scattering (SERS)~\cite{stiles2008surface,acsnano,han2021surface} is one of the most powerful ultra-sensitive optical sensing techniques in which inelastic scattering of light by atoms or molecules is greatly enhanced (by up to a factor of up to $10^{8}$). Prior to the era of SERS, the Raman effect held little appeal for industry due to its weak signal or low scattering cross section~\cite{campion1998surface,mccreery2005raman}. This changed with the advent of surface enhancement, which is accomplished by depositing atoms (or molecules) on a corrugated metal surface or placing them in close proximity to a metallic nanostructure. SERS is distinguished by its ability to accentuate the rich vibrational spectroscopy of an atomic system and promises ample application to electrochemistry~\cite{abdelsalam2005electrochemical}, catalysts~\cite{hartman2016surface}, medicine~\cite{medical_app}, biology~\cite{jamieson2017bioanalytical}, materials science~\cite{SHARMA201216} and others.

SERS was discovered accidentally by Fleischmann and co-workers in 1974 while studying Raman scattering of pyridine on rough silver electrodes~\cite{mcquillan2009discovery}. These results were confirmed by Jeanmaire and Van Duyne in 1977~\cite{jeanmaire1977surface}, and though enhancement factors of $10^{5}$-$10^{6}$ were reported, the underlying physics remained unclear. Subsequent theoretical investigations revealed the enhancement results from localized plasmon excitation on the surface of metallic nanostructures~\cite{lee2007surface,ding2017electromagnetic}. The surface plasmon excitation intensifies the local electromagnetic (EM) field near the placed atoms, enabling a sharp rise in the Raman scattering cross-section as an EM resonance effect. Another widely accepted enhancement mechanism is chemical in nature. It works by charge transfer between molecule and substrate, resulting in an increased polarizability of the molecule. The amplification exhibited by this technique (on the order of $10^{2}$) pales in comparison to its EM counterpart.

The discovery of graphene-enhanced Raman scattering---based on  the graphene or graphene-based substrate---has further ignited research interests in SERS effects~\cite{schedin2010surface}. Although the smooth surface and high optical transmission of graphene do not favor the local EM field enhancement via chemical mechanism, graphene improves the EM near-field sensing SERS performance by imparting a number of synergistic effects, greatly expanding arenas of application~\cite{xu2012surface}. More about graphene-enhanced Raman scattering can be found in Ref.~\cite{lai2018recent}. The single-wall carbon nanotube (SWCN) is formed by rolling up a graphene monolayer into a hollow cylinder~\cite{Saito}. This prototypical one-dimensional (1D) electronic system was previously examined by Bondarev~\cite{bondarev2015plasmon} in the context of SERS as an EM near-field effect due the coupling of a two-level atomic system (TLS) to intrinsic interband plasmon excitations of individual SWCNs, described in a detailed theoretical analysis ensconced in quantum electrodynamics (QED). Prior to this study, the primary means for increasing Raman scattering with CNs was to decorate them with metallic nanoparticles and thereby amplify the EM field through localized plasmon excitation~\cite{Pendry98}.

CNs are extremely flexible and stable materials with externally controllable EM properties. In the last few decades, advancements in fabrication technology have made it possible to significantly improve the quality of SWCNs. This has kindled further investigation over a diverse range of CN applications: single-photon sources~\cite{Optical_Mater_1576,Natute_commu4439,ACS_Nano_11742,Nature_chem1089,Nature_photonics727,Nature_Nanotech671}, field-effect transistors~\cite{Science850,ScienceAdavance_1240}, rechargeable batteries~\cite{Functional_material5018}, controllable
electrical and thermal transport~\cite{Nano_Letter5832}, supercapacitors~\cite{Adv_Energy1814}, and solar cells~\cite{small1150}, to name a few. Recently, the spotlight has been on ultrathin films made of arrays of periodically aligned single-wall carbon nanotubes~\cite{ChemP132,Nature633,Kono2019,NanoLett641,PNAS115,NanoLett19,ACS1602,NanoLett3131,Vac015,NanoLet20,PhyRevApp006,WGao2023,DipJariwala,BondPRappl2021,AdhBondJAL2021,BondCPC2023,Pablo,Pugh}. This system possesses a rough surface that contributes to diminished optical transmissibility (compared with graphene) and might therefore be a better choice for SERS. Such periodically aligned SWCN films extend the study of CN phenomena to the 2D realm.

In general, ultrathin films may be comprised of metals, doped-semiconductors or polar materials and are known to exhibit plasmon-, exciton-, and phonon-polariton eigenmodes~\cite{Nat_Commu1762,ACS_Nano7771,Nature_photon899,Nature_photon216,Nano_Lett984,ACS_Photon2816,PRB121408,BondMousShal2020}. In this context, plasmons become sensitive to the thickness of the film, providing a mechanism for controlled light confinement~\cite{PRB121408,BondShal2017,BondMousShal2020,MRSC2018,Bond2019OMEX}. By reducing the thickness one enhances the inter-electron repulsive potential~\cite{Keldysh,Rytova} as electrons confined in the film interior start interacting more effectively via (lower permittivity) sub- and superstrates, in which case the film's in-plane plasma oscillation frequency becomes nonlocal (spatially dispersive)~\cite{BondShal2017}, the vertical out-of-plane structurization of the film ceases to play its role, and the in-plane structurization remains the only one that matters~\cite{MRSC2018,Bond2019OMEX,BondShal2017,Keldysh}. Recently, it was proposed that such films provide a new regime---transdimensional (TD), in between 3D and 2D, turning into 2D as the film thickness tends to zero~\cite{BoltShal2019,Shah2022}. The remarkable opportunities for tuning their highly confined in-plane plasma modes open access to novel quantum, nonlocal and nonlinear EM effects~\cite{Rivera,BondAnnPhys2023,NonlocRoadmap,BoltRoadmap2025,Zundel,BiehsBond2023,Salihoglu2023,BiehsBond2025,Das2024,Crystalliz2026,Taiwan} in both near-infrared~\cite{BiehsBond2023,Salihoglu2023} and visible ranges~\cite{BiehsBond2025,Taiwan}. Plane-parallel periodically aligned SWCN arrays present a particular case of such TD film systems. Their collective excitations and optical properties have been investigated and shown to be totally in-plane anisotropic~\cite{NanoLett3131,BondPRappl2021}, exhibiting metallic and dielectric in-plane response in the direction of and perpendicular to the CN alignment, respectively~\cite{BondCPC2023,Pablo,Pugh}. Recent ellipsometry measurements~\cite{NanoLett3131,DipJariwala} on such arrays demonstrate a tunable negative dielectric response in the SWCN alignment direction for a wide range of photon excitation energies, in agreement with theoretical predictions~\cite{BondPRappl2021}, giving the promise of highly anisotropic hyperbolic metasurface (MS) development. Hyperbolic metasurfaces have numerous potential applications, ranging from optical sensing, absorption, and cloaking to super-resolution imaging~\cite{Science2013,ChinaTutorial}. SWCN metasurfaces offer tunable structural parameters such as CN diameter, inter-tube spacing and packing in addition to thickness, as well as intrinsic quasiparticle excitation tunability~\cite{DipJariwala,Vasya2025}, which make them flexible multifunctional nanomaterial platforms for even wider scope of applications to include single molecule detection, in particular~\cite{Pugh}.

\begin{figure}[t]
	\includegraphics[width=\textwidth]{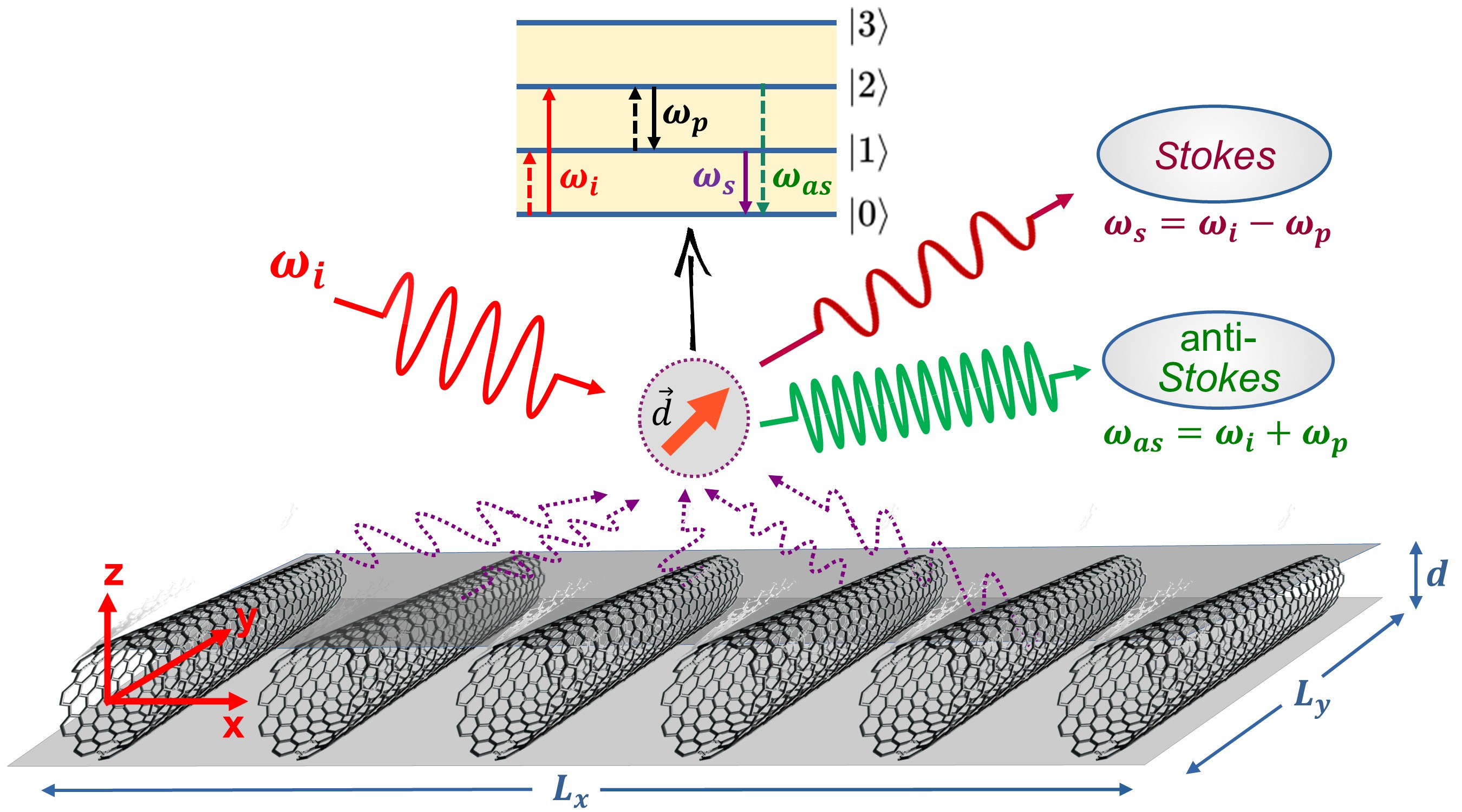}
	\caption{A schematic of the Raman scattering process by a TLS in near proximity to a periodic SWCN array film. The TLS energy spectrum is split into four levels due to the near-field coupling to SWCN plasmon excitations of frequency $\omega_p$ propagating in the CN alignment direction. An incident EM wave of frequency $\omega_{\mts i}$ is scattered inelastically into Stokes and anti-Stokes components of frequencies $\omega_{s}$ and $\omega_{as}$, respectively. Also shown is the four level transition frequency spectrum of the coupled TLS-SWCN film system scattering the external EM wave, with processes involving emission and absorption of a plasmon indicated by the solid and dashed vertical lines, respectively.}
	\label{fig1}
\end{figure}

In this work, we develop a theory of anisotropic SERS phenomena by using a fully-quantized, medium-assisted QED approach for an atom (ion, or molecule) modelled by a TLS placed in near proximity to an ultrathin, closely packed, periodically aligned SWCN film (Fig.~\ref{fig1}). Periodically aligned SWCN films produce strongly anisotropic excitation and emission spectra due to the in-plane linear electromagnetic (EM) response being a tensor with maximal component in the CN alignment direction~\cite{BondPRappl2021,Pugh}, making the respective near-field local photonic density of states (LDOS) orders of magnitude greater than that in the transverse direction. Because of the circumferential quantization of the longitudinal electron motion in individual SWCNs, the real parts of SWCN axial optical conductivities (or imaginary parts of the EM response functions along the SWCN symmetry axis, equivalently) consist of the 1st, 2nd, 3rd, etc., exciton resonances~\cite{Ando}. The imaginary parts of the axial conductivities are linked with real ones by Kramers-Kr\"{o}nig relation, and so the real parts of the \emph{inverse} axial conductivities (energy loss functions given equivalently by imaginary parts of inverse EM response functions along the SWCN axis) show the resonances next to their excitonic counterparts as well. These are interband plasmons---\emph{standing} charge density waves due to the periodic opposite-phase displacements of electron shells with respect to ion cores in individual SWCNs~\cite{Bondarev12,Bondarev12pss,Bondarev14}. Excitons and interband plasmons originate from the same circumferentially quantized electronic transitions~\cite{Bondarev09} (though excited differently---by transversely and longitudinally polarized EM radiation), producing photonic LDOS resonances in the far- and near-field zone, respectively, and giving rise to series of quantum negative refraction bands, accordingly~\cite{BondPRappl2021}. Experimental evidence for low-energy ($\sim\!1\!-\!2$~eV) interband plasma modes in CNs was first reported by Pichler et al.~\cite{Pichler98}. When excited, their near-surface quasistatic electric fields can be strong enough to result in the enhanced Raman scattering effect by atomic type species (extrinsic atoms, ions, molecules, or semiconductor quantum dots) in the CN vicinity~\cite{bondarev2015plasmon}. The effect remains strong (at least for the lowest negative refraction band) in periodically aligned SWCN films as well~\cite{Pugh,DipJariwala}.

This work extends and further develops the electromagnetic SERS effect theory previously reported by one of us for a TLS near a single-wall carbon nanotube~\cite{bondarev2015plasmon}. Among other things, an obvious nontrivial difference now is a realistic quasi-2D quantum material: an ultrathin periodically aligned SWCN film to provide an in-plane anisotropic interface with collective band structure, collective intrinsic quasiparticle excitations (excitons and plasmons), and a well-defined incidence plane for incoming and scattered light radiation, which can yet be oriented arbitrarily relative to the main anisotropy (CN alignement) axis. We derive and analyze the differential scattering cross-section for the TLS coupled to the plasmon-mediated quasistatic electric fields in the near-field zone of such an ultrathin anisotropic MS. Particular attention is given to the polarization of external light radiation. By explicitly deriving the polarization tensor of scattered light, we establish that for the TLS located in the near-field zone of the periodically aligned SWCN metasurface its Raman cross-section components (Stokes and anti-Stokes) can be enhanced by a factor of up to $10^{4}$, not only for $p$-polarized but also for $s$-polarized incoming light radiation.

\begin{figure}[t]
	\includegraphics[width=\textwidth]{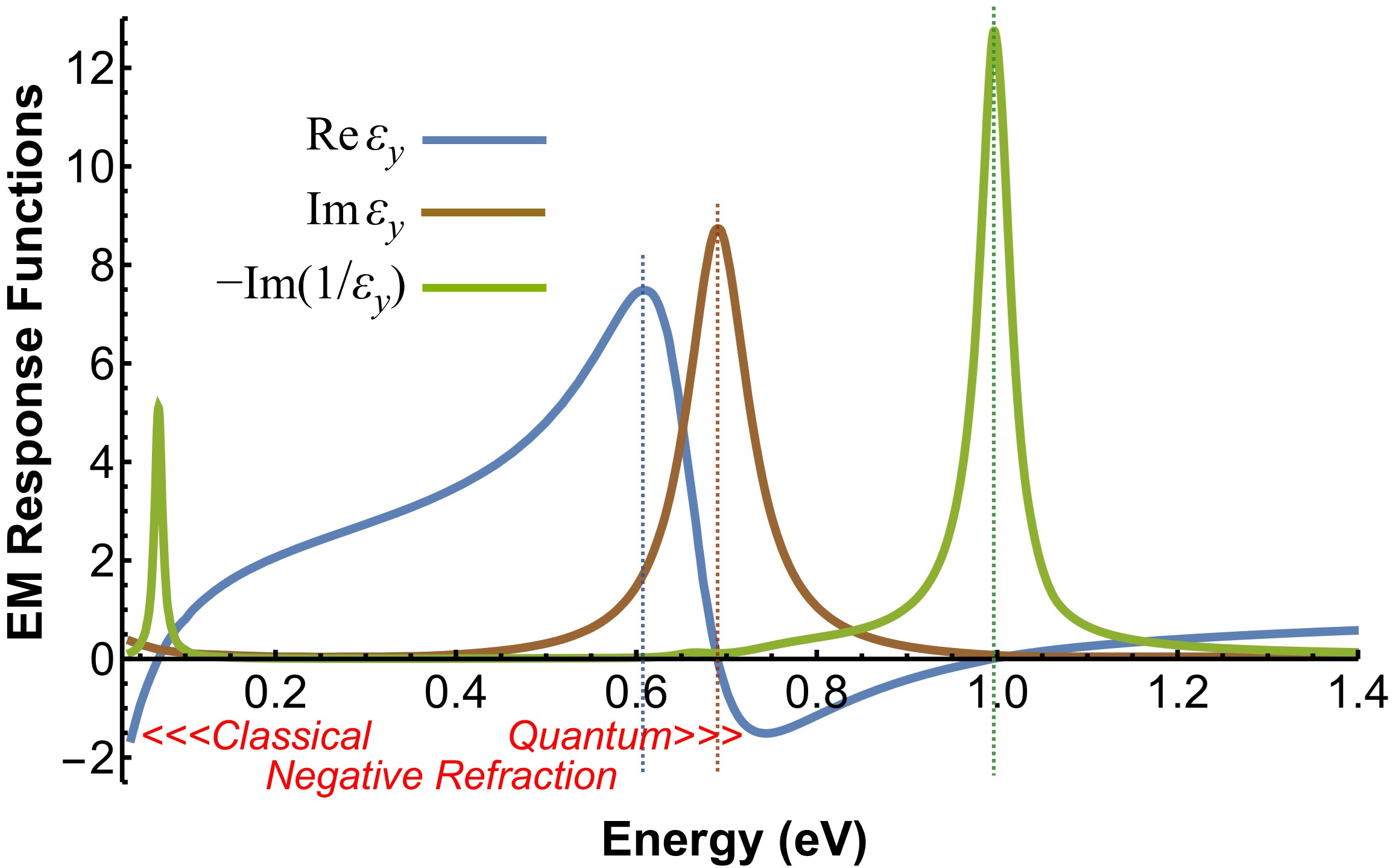}
	\caption{Real (blue) and imaginary (brown) parts of the low/medium excitation energy in-plane EM response tensor  along SWCN alignment direction ($y$-direction in Fig.~\ref{fig1}) used to model the linear EM response for ultrathin semiconducting SWCN films~\cite{Pugh}. The energy loss function shown by the green line exhibits classical (left) and \textit{quantum} interband (right) plasmon resonances associated with the classical and \textit{quantum} negative refraction bands due to the Kramers-Kr\"{o}nig relations (see Ref.~\cite{Pugh} for details).}
	\label{fig2}
\end{figure}

\section{Medium-Assisted QED Formalism}

The problem of the polarized light radiation scattering by an atom in close proximity to a metasurface requires the exact knowledge of the quantized energy spectrum of the coupled atom-MS system. Our MS (sketched in Fig.~\ref{fig1}) is formed by a horizontal periodic (quasi-periodic) homogeneous (quasi-inhomogeneous) array of semiconducting SWCNs aligned in the $y$-direction, with its translational unit in the $x$-direction and thickness $\sfit{d}$ being less than the wavelength of the external radiation (a condition necessary for the dipole approximation used below~\cite{Pugh}). The array is assumed to be embedded in a solid dielectric to form a composite layer with in-plane effective static dielectric permittivity $\varepsilon$ being greater than $\epsilon_1$ and $\epsilon_2$, the static permittivities of the substrate and superstrate, respectively. The electron charge density of such a system is constrained by the SWCN cylindrical symmetry to be uniformly and periodically distributed over SWCN surfaces. The positive background of nuclei keeps the entire system electrically neutral. The collective linear in-plane EM response in the direction of alignment for such an MS system was previously derived and analyzed theoretically by one of us~\cite{BondPRappl2021}, and was shown to be of the general form presented in Fig.~\ref{fig2} in the low and medium excitation energy domain containing the 1st exciton and interband plasmon resonances, in full agreement with experiments reported before~\cite{NanoLett3131,Vac015}. The theory used the finite-temperature Matsubara Green's function formalism to include \textit{all} quantized quasiparticle excitations, which led to the prediction of quantum negative refraction bands (Fig.~\ref{fig2}) discovered in another set of experiments very recently~\cite{DipJariwala}. More specifically, the EM response shown in Fig.~\ref{fig2} features the lowest exciton absorption resonance (peak of $\mbox{Im}\,\varepsilon_y$, brown line) as well as classical (pertinent to bulk metals) and quantum (originating from the quantum confined electron motion on the SWCN surface) interband plasmon resonances of the function $-\mbox{Im}\,(1/\varepsilon_y)$ (energy loss-function, green line, lower and higher energy peaks, respectively) associated with zeros of $\mbox{Re}\,\varepsilon_y$ (optical refraction function, blue line) according to the Kramers-Kr\"{o}nig relations. It was also shown that with $\sfit{d}$ decreasing and becoming less than the mean in-plane distance between a pair of electrons in the composite layer, the ultrathin SWCN metasurface enters the TD regime where its dimensionality is reduced from 3D to 2D and its vertical out-of-plane structurization is only controlled by the CN volumetric fraction (not by CN vertical ordering), which includes layer thickness $\sfit{d}$ as the only remnant of the $z$-direction.

In the medium-assisted QED formalism we use here to derive the Raman scattering cross-section by the coupled atom-MS system, intrinsic quasiparticle (phonon, plasmon, exciton, etc.) relaxation phenomena started by external light radiation are considered to create random field fluctuations superimposed on those of the physical vacuum where no medium is present~\cite{Pugh,WelschQO}. These medium-assisted fluctuating vacuum-type EM fields can be represented by the quantum electric and magnetic field operators (Schrödinger picture, Gaussian units)
\begin{eqnarray}
	\bwhat{\vc E}(\vc r)=\bwhat{\vc E}^{(+)}(\vc r)+\bwhat{\vc E}^{(-)}(\vc r)\;\;\;\mbox{and}\;\;\;\bwhat{\vc H}(\vc r)=-\ii\frac{c}{\omega}\,\vc\nabla\times\bwhat{\vc E}(\vc r),\label{QFields}\\
	\bwhat{\vc E}^{(+)}(\vc r)=\Int 0{\infty}\!\d{\omega}\ub{\bwhat{\vc E}}(\vc r,\omega),\;\;\;\bwhat{\vc E}^{(-)}(\vc r)=[\bwhat{\vc E}^{(+)}(\vc r)]^{\dagger}.\hskip0.75cm\nonumber
\end{eqnarray}
Here, $\vc r=\vc{\rho}+z\uv z$ where $\vc{\rho}$ is in plane and $\uv z$ is perpendicular to plane of the MS, respectively. The Fourier image components of $\bwhat{\vc E}$ in directions $\alpha=x,y,z$ are given by
\begin{equation}
	\ub{\bwhat E}_{\alpha}(\vc r,\omega)=\ii\frac{4\pp\omega}{c^{2}}\!\int\!\d{\vc{\rho}}\sum_{\mathclap{\lambda=x,y,z}}G_{\alpha\lambda}(\vc r,\vc{\rho},\omega)\;\bwhat{\!\!\ub{\,J}}_{\lambda}(\vc{\rho},\omega),
	\label{QFieldE}
\end{equation}
where
\begin{equation}
	\bwhat{\!\!\ub{\,J}}_{\alpha}(\vc{\rho},\omega)=\frac{\omega}{2\pp}\sqrt{\hbar \sfit{d}\Im\varepsilon_{\alpha\alpha}(\vc{\rho},\omega)}\,\bwhat f_{\alpha}(\vc{\rho},\omega)\label{QFieldJ}
\end{equation}
is the quantum noise current density operator to warrant the correct equal-time electric-magnetic field operator commutation relations of standard vacuum QED and Fluctuation-Dissipation Theorem fulfilment in the presence of medium~\cite{WelschQO}. Here, $G_{\alpha\lambda}$ and $\varepsilon_{\alpha\alpha}$ are the components of Green's and diagonalized linear EM response tensors responsible for geometry and material EM properties of the MS system, respectively, and $\bwhat f_{\alpha}$ are the annihilation operators for material intrinsic bosonic quasiparticle excitations which commute with their creation counterparts via the rule
\[
[\bwhat f_{\alpha}(\vc{\rho},\omega),\bwhat f^{\;\dagger}_{\!\beta}(\vc{\rho}',\omega')]=\kd{\alpha\beta\,}\kd{}(\vc{\rho}-\vc{\rho}')\,\kd{}(\omega-\omega').
\]

In terms of the medium-assisted QED scheme above, the following second-quantized Hamiltonian describes the coupled atom-field system with an atom (or a molecule) modeled by the TLS positioned at an arbitrary point $\vc r_{A}=\vc{\rho}_A+z_A\vc{e}_z$ above the SWCN film (as sketched in Fig.~\ref{fig1}) within the framework of the electric dipole and two-level approximations~\cite{WelschQO,BondLamb06,BondLamb05,BondLamb04}
\begin{eqnarray}
	\bwhat{H}\!&=&\!\bwhat{H}_F+\bwhat{H}_A+\bwhat{H}_{AF}=\bwhat{H}_F+\bwhat{H}_A+\bwhat{H}^{(a)}_{AF}+\bwhat{H}^{(e)}_{AF}\label{Htwolev}\\
	&=&\!\!\!\Int 0{\infty}\!\d{\omega}\!\int\!\d{\vc{\rho}}\sum_{\lambda}\hbar\omega\bwhat f^{\,\dagger}_{\lambda}(\vc{\rho},\omega)\bwhat f_{\lambda}(\vc{\rho},\omega)+\frac{1}{2}\hbar\omega_{A}\bwhat{\sigma}_{z}-\sum_{\lambda}\big[d_{\lambda}\bwhat E^{{\scriptscriptstyle (+)}}_{\lambda}(\vc r_{A})\bwhat{\sigma}^{\dagger}\!+d_{\lambda}^\ast\bwhat E^{{\scriptscriptstyle (-)}}_{\lambda}(\vc r_{A})\bwhat{\sigma}\big]\nonumber
\end{eqnarray}
Here, the three terms represent the MS quantum field subsystem, the atomic subsystem modeled by a two-level dipole emitter with transition frequency $\omega_A$, and their interaction composed of absorption and emission terms for material single-quantum bosonic excitations, respectively. The Pauli operators $\bwhat{\sigma}_{z}\!=\!|u\rangle\langle u|\!-\!|l\rangle\langle l|$, $\bwhat{\sigma}\!=\!|l\rangle\langle u|$, $\bwhat{\sigma}^{\dag}\!=\!|u\rangle\langle l|$ describe the transitions in the atomic subsystem between its upper $|u\rangle$ and lower $|l\rangle$ states. The interaction term is due to the coupling of the atomic transition dipole $d_\lambda\!=\!\langle u|\bwhat{d}_\lambda|l\rangle$ ($\lambda\!=\!x,y,z$) to the quantum medium-assisted electric field of Eqs.~(\ref{QFields})--(\ref{QFieldJ}). The frequency $\omega_{A}$ is generally red-shifted relative to the transition dipole frequency $\omega_{ul}$, so that $\omega_A\!=\omega_{ul}+\delta\omega_{ul}$ ($\delta\omega_{ul}\!<\!0$), due to the van der Waals coupling of the dipole emitter to the film~\cite{BondLamb06,BondLamb05}. For simplicity and with no loss of generality, the shift $\delta\omega_{ul}$ will be assumed being included in $\omega_A$ here. Its general properties can be found in Ref.~\cite{WelschQO}.

In the linear coupling regime, quite generally, the coupled MS--TLS system can be represented as a four-level system (sketched in Fig.~\ref{fig1}) with eigenvectors of the Hamiltonian (\ref{Htwolev}) of the form
\begin{eqnarray}
	|0\rangle\!\!&=&\!\!|l\rangle|\{0\}\rangle,\nonumber\\
	|1,2\rangle\!\!&=&\!\!C_u^{(1,2)}|u\rangle|\{0\}\rangle+\!\int_{0}^{\infty}\!\!\!\!\!\!\d\omega\!\int\!\!\d{\vc{\rho}}\sum_{\lambda}C^{(1,2)}_{l\lambda}(\vc{\rho},\omega)\,|l\rangle|\{1_\lambda({\vc{\rho}},\omega)\}\rangle,\label{4states}\\
	|3\rangle\!\!&=&\!\!|u\rangle|\{1_\lambda({\vc{\rho}},\omega)\}\rangle\nonumber,
\end{eqnarray}
with normalization condition
\begin{equation}
	|C^{(1,2)}_{u}|^{2}+\Int 0{\infty}\!\d{\omega}\!\int\!\d{\vc{\rho}}\sum_{\lambda}|C^{(1,2)}_{l\lambda}(\vc{\rho},\omega)|^{2}=1.
	\label{norm}
\end{equation}
Here, $|\{0\}\rangle$ and $|\{1_\lambda({\vc{\rho}},\omega)\}\rangle\!=\!\bwhat f_{\lambda}^{\,\dagger}(\vc{\rho},\omega)|\{0\}\rangle$ are the ground and single-quantum excited Fock states, respectively, of the SWCN metasurface subsystem, and $C_{u,l}^{(1,2)}$ are unknown mixing coefficients for non-radiative virtual spontaneous transition $|u\rangle|\{0\}\rangle\!\rightarrow\!|l\rangle|\{1_\lambda({\vc{\rho}},\omega)\}\rangle$. The transition between these quasi-degenerate quantum states of the coupled TLS--MS system generates single-quantum bosonic excitations of frequency $\omega$ at point $\vc{\rho}$ of the MS with simultaneous TLS de-excitation. The mixing coefficients can be found by solving the eigenvalue problem for Hamiltonian (\ref{Htwolev}) in basis~(\ref{4states}) in the same manner as it was earlier done by one of us for the single SWCN case~\cite{bondarev2015plasmon,BondLamb06,BondLamb05}.

Proceeding as in Ref.~\cite{bondarev2015plasmon}, one obtains
\begin{equation}
	E_{0}=-\frac{\hbar\omega_{A}}{2}\;\;\;\text{and}\;\;\;E_{3}=\frac{\hbar\omega_{A}}{2}+\hbar\omega\label{E03}
\end{equation}
for eigenvectors $|0\rangle$ and $|3\rangle$, respectively, and a closed simultaneous set of two algebraic equations (two-state eigenvalue problem) for the mixing coefficients and energy eigenvalues $E_{1}$ and $E_{2}$ of eigenvectors $|1\rangle$ and $|2\rangle$, which in the Green's tensor principle axis system yields
\begin{equation}
	C^{(1,2)}_{l\lambda}(\vc{\rho},\omega)=\ii\frac{2\omega^{2}}{c^{2}}\frac{\sqrt{\hbar \sfit{d}\Im\varepsilon_{\lambda\lambda}(\vc{\rho},\omega)}}{\hbar\omega_{A}/2-\hbar\omega+E}\sum_{\mu}d_{\mu}^\ast G^{\ast}_{\mu\lambda}(\vc r_{A},\vc{\rho},\omega)\,C^{(1,2)}_{u}.\label{Cl}
\end{equation}
This, after using the normalization condition above and general identity
\begin{equation}
	\frac{\omega^{2}}{c^{2}}\!\int\!\!\d{\vc{\rho}}\sum_{\lambda}\sfit{d}\Im\varepsilon_{\lambda\lambda}(\vc{\rho},\omega)\,G_{\alpha\lambda}(\vc r,\vc{\rho},\omega)G^{\ast}_{\beta\lambda}(\vc{r}^\prime\!,\vc{\rho},\omega)=\Im G_{\alpha\beta}(\vc r,\vc{r}^\prime\!,\omega),
	\label{identity}
\end{equation}
($\sfit{d}$ is the MS thickness, see Refs.~\cite{WelschQO,Pugh}), leads to
\begin{equation}
	C^{(1,2)}_{u}=\Bigg[1-\frac{4\hbar}{c^2}\sum_{\mu,\nu}d_{\mu}d_{\nu}^\ast\!\Int 0{\infty}\!\d{\omega}\omega^{2}\frac{\Im G_{\mu\nu}(\vc r_{A},\vc r_{A},\omega)}{(\hbar\omega_{A}/2-\hbar\omega+E)^2}\Bigg]^{-1/2}\!\!\!,
	\label{Cu}
\end{equation}
whereby the two-state eigenvalue problem returns the following integral equation
\begin{equation}
	E=\frac{\hbar\omega_{A}}{2}+\frac{4\hbar}{c^{2}}\sum_{\mu,\nu}d_{\mu}d_{\nu}^\ast\!\Int 0{\infty}\!\d{\omega}\omega^{2}\frac{\Im G_{\mu\nu}(\vc r_{A},\vc r_{A},\omega)}{\hbar\omega_{A}/2-\hbar\omega+E}\label{Eint}
\end{equation}
to be solved for the energy eigenvalues $E_{1,2}$ of the eigenvectors $|1,2\rangle$. Plugging the solutions of Eq.~(\ref{Eint}) in Eqs.~(\ref{Cl}) and (\ref{Cu}) completes the two-state eigenvalue problem solving for our coupled MS--TLS system.

\section{Raman Scattering Cross-Section}

\begin{figure}[t]
	\begin{centering}
		\includegraphics[width=\textwidth]{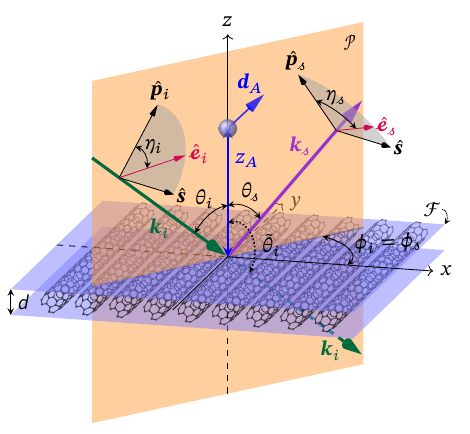}
		\par\end{centering}
	\caption{TLS--SWCN film system schematic showing incident $\vc k_{\mts i}$ and scattered $\vc k_{\mts s}$ photon momenta along with their respective orthogonal polarization vectors $\hat{\mathbfscr{e}}_{\mts i}$ and 	 $\hat{\mts{\mathbfscr{e}}}_{\mts s}$. The SWCN film $\mathcal F$ makes the ($x,y$)-plane with SWCN alignment parallel to the $y$-axis. The plane of incidence $\mathcal P$ passes through $z$-axis and makes angle $\phi_{\mts i}\!=\phi_{\mts s}$ with $x$-axis. The angle of incidence is $\theta_{\mts i}\!=\theta_{\mts s}$. The angle $\eta_{\mts i(\mts s)}$ is that $\hat{\mathbfscr{e}}_{\mts i(\mts s)}$ makes with unit vector $\hat{\vc p}_{\mts i(\mts s)}$, which lies at the intersection of plane $\mathcal P$ and a plane transverse to $\vc k_{\mts i(\mts s)}$. When $\eta_{\mts i(\mts s)}=0$, $\hat{\mathbfscr{e}}_{\mts i(\mts s)}$ is parallel to $\hat{\vc p}_{\mts i(\mts s)}$ and the photon is $p$-polarized; when $\eta_{\mts i(\mts s)}=\pp/2$, $\hat{\mathbfscr{e}}_{\mts i(\mts s)}$ is parallel to unit vector $\hat{\vc s}$ orthogonal to plane $\mathcal P$ and the photon is $s$-polarized.}
	\label{fig3}
\end{figure}

Under the assumption that the coupled MS--TLS system with eigen states (\ref{4states})--(\ref{Eint}) is initially in the ground state, the inelastic scattering of external EM radiation by this system only involves transitions between levels $|0\rangle$, $|1\rangle$ and $|2\rangle$ as shown in Fig.~\ref{fig1}. The entire scattering process is illustrated in Fig.~\ref{fig3}. The process includes the following three sequential steps~\cite{bondarev2015plasmon,Cardona}:

(a)~excitation of the MS--TLS system by an incident photon of frequency $\omega_{\mts i}$ and unit polarization vector $\uvcr{i}$, described by the interaction matrix element
\begin{equation}
	\langle n|\bwhat{H}_R(\omega_{\mts i})|0\rangle=-\frac{i}{c}\sqrt{2\pp\hbar\omega_{\mts i}}\,C_u^{(n)\ast}\sum_\lambda
	d_{\lambda\,}\uscr i\lambda\,,\;\;\;n\!=\!1,2
	\label{HR}
\end{equation}
(normalized at one photon per unit volume~\cite{Berest});

(b)~emission (or absorption) of a single-quantum MS excitation with matrix element
\begin{equation}
	\langle 1|\bwhat{H}^{(e)}_{AF}|2\rangle=\ii\!\!\int_{0}^{\infty}\!\!\!\!\!\d\omega\frac{2\omega^2}{c^2}\!\!\int\!\!\d{\vc{\rho}}\sum_{\lambda,\mu}C_{l\lambda}^{(1)\ast}(\vc{\rho},\omega)\sqrt{\hbar\sfit{d}\Im\varepsilon_{\lambda\lambda}(\vc{\rho},\omega)}\,d_{\mu}^\ast G^{\ast}_{\mu\lambda}(\vc r_{A},\vc{\rho},\omega)\,C_u^{(2)}
	\label{HAF}
\end{equation}
(or $\langle 2|\bwhat{H}^{(a)}_{AF}|1\rangle=\langle 1|\bwhat{H}^{(e)}_{AF}|2\rangle^\dagger$ for absorption), where $\bwhat{H}^{(e)}_{AF}$ and $\bwhat{H}^{(a)}_{AF}$ are given by the respective parts of the interaction Hamiltonian $\bwhat{H}_{AF}$ in Eq.~(\ref{Htwolev});

(c)~de-excitation of~the MS--TLS system by means of the scattered (Raman) photon emission of frequency $\omega_{\mts s}$ with unit polarization vector $\uvcr{s}$, represented by the matrix element $\langle n|\bwhat{H}_R(\omega_{\mts i})|0\rangle^\dagger|_{\mts{i}\to\mts{s}}$ in accordance with Eq.~(\ref{HR}).

The Fermi Golden Rule transition rate for the process described is given by the sum of emission and absorption terms as follows~\cite{bondarev2015plasmon,Cardona}
\begin{equation}
	\ocurl T=\frac{2\pp}{\hbar}\left[\biggl|\frac{\nwcal N_{\mts{si}}}{\nwcal D^{-}_{\mts{si}}}+\frac{\nwcal N_{\mts{is}}}{\nwcal D^{+}_{\mts{is}}}\biggr|^{2}\!\kd{}(\hbar\omega_{\mts i}-\hbar\omega_{p}-\hbar\omega_{\mts s})+\biggl|\frac{\nwcal N^{\ast}_{\mts{si}}}{\nwcal D^{-}_{\mts{is}}}+\frac{\nwcal N^{\ast}_{\mts{is}}}{\nwcal D^{+}_{\mts{si}}}\biggr|^{2}\!\kd{}(\hbar\omega_{\mts i}+\hbar\omega_{p}-\hbar\omega_{\mts s})\right],
\end{equation}
where the delta-functions stress total energy conservation during photon scattering with absorption or emission of an MS excitation quantum $\hbar\omega_p$ without change of the state of the MS--TLS system,
\begin{align*}
	\nwcal N_{\mts{si}} & =\langle0\vert\bwhat H_{R}(\omega_{\mts s})\vert1\rangle\langle1\vert\bwhat H^{(e)}_{AF}\vert2\rangle\langle2\vert\bwhat H_{R}(\omega_{\mts i})\vert0\rangle
\end{align*}
and
\begin{align*}
	\nwcal D^{\pm}_{\mts{si}} & =[\hbar\omega_{\mts s}\pm(E_{1}-E_{0})][\hbar\omega_{\mts i}\pm(E_{2}-E_{0})]
\end{align*}
with $E_{1,2}$ given by solutions to Eq.~(\ref{Eint}) above. Here, it can be seen that $\nwcal D^{+}_{\mts{is}}$ and $\nwcal D^{+}_{\mts{si}}$ can never be zero, meaning that the terms containing them are \emph{off-resonance}, can only contribute to the background, and so can be neglected. Using Eqs.~(\ref{HR}), (\ref{HAF}), (\ref{Cl}) and (\ref{identity}) in what remains, one arrives at
\begin{eqnarray}
	\ocurl T&\!\!\!=\!\!\!&\frac{2\pp}{\hbar}\biggl[\frac{8\pp\hbar^2}{c^{4}}|C^{(2)}_{u}|^{2}|C^{(1)}_{u}|^{2}\biggr]^{2}\omega_{\mts i}\omega_{\mts s}\,\biggl|\bsum_{\lambda,\mu,\nu,\eta}\!\!\!d_{\lambda}d^{\ast}_{\mu}d_{\nu}d^{\ast}_{\eta}\uscr s{\lambda}^{\ast}\uscr i{\mu}\mts G_{\nu\eta}(\vc r_{A})\biggr|^{2}\label{eq:TransitionRate}\\
	&\!\!\!\times\!\!\!&\left\{\frac{\kd{}(\hbar\omega_{\mts i}-\hbar\omega_{p}-\hbar\omega_{\mts s})}{\bigl|{\nwcal D^{-}_{\mts{si}}}\bigr|^{2}}+\frac{\kd{}(\hbar\omega_{\mts i}+\hbar\omega_{p}-\hbar\omega_{\mts s})}{\bigl|{\nwcal D^{-}_{\mts{is}}}\bigr|^{2}}\right\},\nonumber
\end{eqnarray}
where
\begin{equation}
	\mts G_{\nu\eta}(\vc r_{A})=\!\!\Int 0{\infty}\!\!\d{\omega}\omega^{2}\frac{\Im G_{\nu\eta}(\vc r_{A},\vc r_{A},\omega)}{\hbar\omega_{A}/2-\hbar\omega+E_{1}}\,.
	\label{eq:GreenIntegral}
\end{equation}

To obtain the differential scattering cross-section, one has to average Eq.~(\ref{eq:TransitionRate}) over initial transition dipole orientations and multiply it by the density of final states $(\hbar\omega_{\mts s})^2d\Omega_{\mts s}/(2\pp\hbar)^3$ for photons scattered within the solid angle $d\Omega_{\mts s}$ of the free space~\cite{Berest}. The averaging of the initial dipole orientations leads to
\begin{equation}
	\biggl|\bsum_{\lambda,\mu,\nu,\eta}\!\!\!d_{\lambda}d^{\ast}_{\mu}d_{\nu}d^{\ast}_{\eta}\uscr s{\lambda}^{\ast}\uscr i{\mu}\mts G_{\nu\eta}(\vc r_{A})\,\biggr|^{2}\!=\frac{|d|^8}{3^4}\,\biggl|\,2\bsum_{\lambda,\mu}\uscr s{\lambda}^{\ast}\overline{\mts G}_{\lambda\mu}(\vc r_{A})\uscr i{\mu}+\frac{5}{3}\uvcr s^{\ast}\!\cdot\!\uvcr i\tr\mts G(\vc r_{A})\,\biggr|^{2}
	\label{ddddfinal}
\end{equation}
(see Appendix A), where
\begin{equation}
	\overline{\mts G}_{\lambda\mu}(\vc r_{A})=\frac{1}{2}\,\biggl[\mts G_{\lambda\mu}(\vc r_{A})+\mts G_{\mu\lambda}(\vc r_{A})-\frac{2}{3}\kd{\lambda\mu}\tr\mts G(\vc r_{A})\biggr]\,
	\label{traceless}
\end{equation}
is the traceless symmetric tensor built of the components of that in Eq.~(\ref{eq:GreenIntegral}). For centrosymmetric \textit{achiral} SWCN film systems we are focusing on here, one has $\mts G_{\lambda\mu}=\kd{\lambda\mu}\ocurl{\mts G}_{\lambda\lambda}$ in the Cartesian principle axes coordinates shown in Fig.~\ref{fig1}. (Films of chiral SWCNs will be discussed elsewhere.) Then, $\overline{\ocurl{\mts G}}_{\lambda\mu}=(\ocurl{\mts G}_{\lambda\lambda}-\tr\ocurl{\mts G}/3)\kd{\lambda\mu}$ where, in addition, $\tr\mts G\approx\mts G_{yy}$ as the component in the SWCN alignment direction can be shown to dominate for TLS residing in the near-field zone with $\hbar\omega_A$ being within the MS quantum negative refraction band (see Fig.~\ref{fig2})---the conditions to be fulfilled for the resonance TLS--SWCN coupling we discussed recently~\cite{Pugh}. This allows us to factor out incoming photon's angular variables sketched and defined in Fig.~\ref{fig3}. We have
\begin{equation}
	\biggl|\bsum_{\lambda,\mu,\nu,\eta}\!\!\!d_{\lambda}d^{\ast}_{\mu}d_{\nu}d^{\ast}_{\eta}\uscr s{\lambda}^{\ast}\uscr i{\mu}\mts G_{\nu\eta}(\vc r_{A})\,\biggr|^{2}\!\approx\frac{|d|^8}{3^4}|\mts G_{yy}(\vc r_{A})|^2F(\eta_{\mts i},\phi_{\mts i},\theta_{\mts i}),
\end{equation}
where the angular factor is
\begin{eqnarray}
	F(\eta_{\mts i},\phi_{\mts i},\theta_{\mts i})\!&=&\!\Big|\uscr sx^{\ast}\uscr ix+3\uscr sy^{\ast}\uscr iy+\uscr sz^{\ast}\uscr iz\Big|^{2}\label{AngularFactor}\\
	&=&\!\left|1\!+\!2\cos^{2}\phi_{\mts i}\!-\!2\cos^{2}\eta_{\mts i}\big(\!\cos^{2}\phi_{\mts i}\sin^{2}\theta_{\mts i}\!+\!2\cos^{2}\theta_{\mts i}\!\big)\right|^{2}\nonumber
\end{eqnarray}
(see Appendix B). Finally, with substitutions for $C_u^{(1,2)}$ and $E_{1,2}$ obtained under the same resonance conditions from Eqs.~(\ref{Cu}) and (\ref{Eint}), respectively, as described in Appendix~C, the differential scattering cross-section can be conveniently written in the dimensionless form with angular, TLS--MS coupling strength and energy variables separated as follows
\begin{equation}
	\frac{\d{\sigma}}{\d{\Omega_{\mts s}}}\approx\frac{(2\gamma_0)^2|d|^{4}}{9c^{4}\hbar^{4}}S(x_{\mts i},x_{\mts s},\eta_{\mts i},\phi_{\mts i},\theta_{\mts i}),
	\label{cross-sec}
\end{equation}
where
\begin{equation}
	S(x_{\mts i},x_{\mts s},\eta_{\mts i},\phi_{\mts i},\theta_{\mts i})=F(\eta_{\mts i},\phi_{\mts i},\theta_{\mts i})\,P(x_{\mts i},x_{\mts s})
	\label{S}
\end{equation}
and
\begin{eqnarray}
	P(x_{\mts i},x_{\mts s})\hskip-0.1cm&=&\hskip-0.1cm x_{\mts i}x_{\mts s}^3\,A(\delta,X,\Delta x_p)\left\{\frac{1}{\big[\!(x_{\mts s}\!-\!x_p\!-\!\delta_{-}/2)^2\!+\!(\Delta x_p/2)^2\big]\big[\!(x_{\mts i}\!-\!x_p\!-\!\delta_{+}/2)^2\!+\!(\Delta x_p/2)^2\big]}\right.\nonumber\\
	&+&\hskip-0.1cm\left.\frac{1}{\big[\!(x_{\mts s}\!-\!x_p\!-\!\delta_{+}/2)^2\!+\!(\Delta x_p/2)^2\big]\big[\!(x_{\mts i}\!-\!x_p\!-\!\delta_{-}/2)^2\!+\!(\Delta x_p/2)^2\big]}\right\}
	\label{Pis}
\end{eqnarray}
is the dimensionless angle-free scattering probability function with amplification factor
\begin{equation}
	A(\delta,X,\Delta x_{p})=\frac{X^{8}}{2^{6}(\delta^{2}+X^{2})^{2}[\delta^{2}_{-}+(\Delta x_{p})^{2}\bigr]}.
	\label{A}
\end{equation}
In these equations, $(x_{\mts i,\mts s},x_A,x_p,\Delta x_p)=\hbar(\omega_{\mts i,\mts s},\omega_A,\omega_p,\Delta\omega_p)/(2\gamma_{0})$ are dimensionless energies ($\gamma_0=2.7$~eV is the graphene carbon-carbon overlap integral) representing incident and scattered photons, TLS atomic transition, MS quantum interband plasmon resonance (shown by the vertical dotted green line in Fig.~\ref{fig2} for a typical SWCN MS system) and its half-width-at-half-maximum, respectively; $\delta\!=\!x_A-x_p$, $\delta_{\pm}\!=\!\delta\pm\sqrt{\delta^2+X^2}$, and $X\!=\!\hbar\sqrt{2\Delta\omega_{0}\Gamma_0(\omega_p)\xi_y(\vc{r}_{A},\omega_p)}/(2\gamma_{0})$ is the parameter (Rabi splitting, see Appendix C) that controls the TLS coupling to the MS plasmon mode providing the resonance photon-scattering conditions through the drastic enhancement of the $\xi_y(\vc{r}_{A},\omega_p)$ function, local photonic density-of-states (LDOS), for TLS in the MS evanescent field zone in the manner we recently discussed for resonance fluorescence in the same system~\cite{Pugh}.

\begin{figure}[t]
	\begin{centering}
		\includegraphics[width=\textwidth]{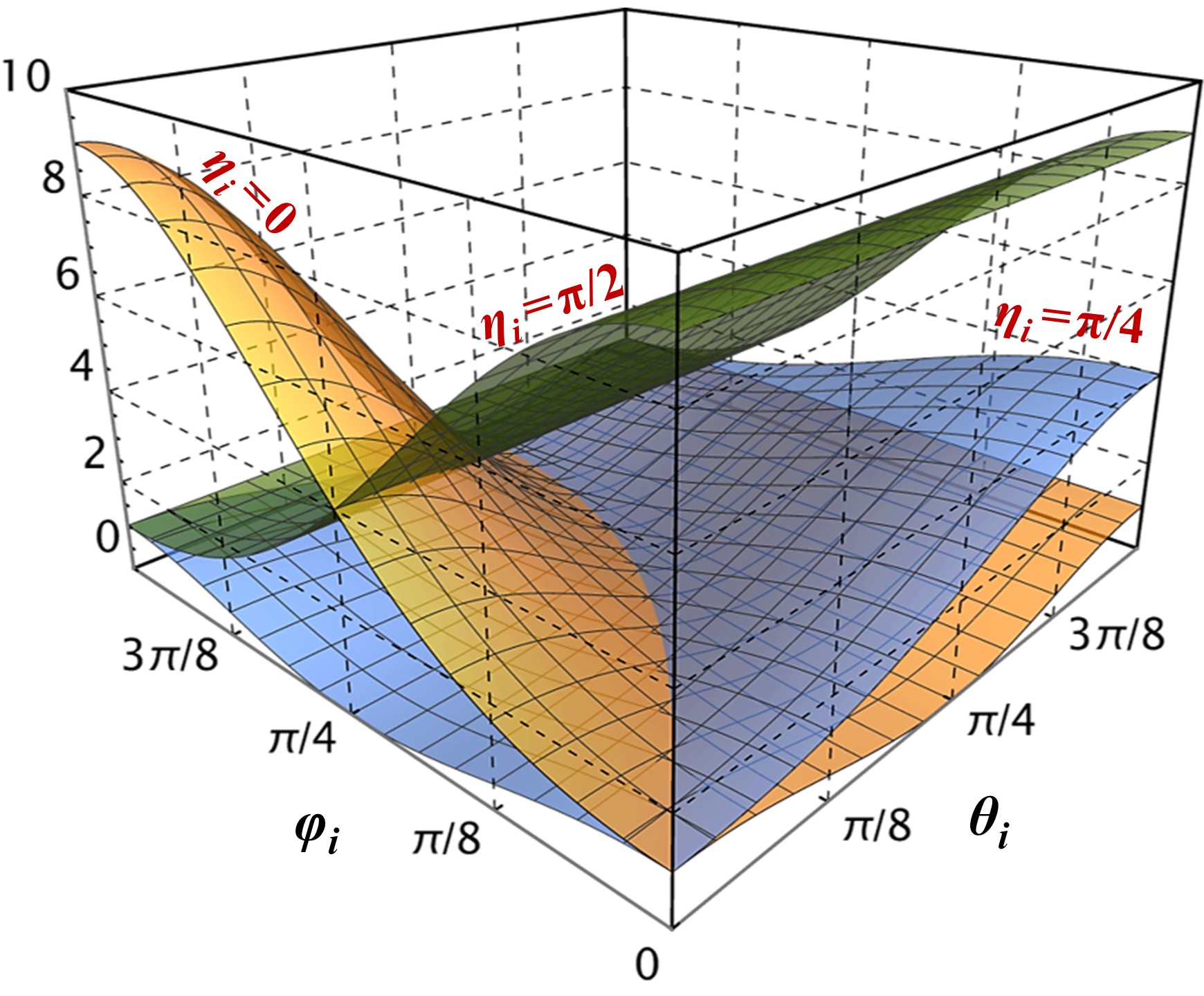}
		\par\end{centering}
	\caption{Angular factor $F(\eta_{\mts i},\phi_{\mts i},\theta_{\mts i})$ of differential scattering cross-section (\ref{cross-sec})--(\ref{A}), as given by Eq.~(\ref{AngularFactor}) for $\eta_{\mts i}\!=0$ ($p$-polarization), $\pp/4$, and $\pp/2$ ($s$-polarization). See Fig.~\ref{fig3} for angle definitions.}
	\label{fig4}
\end{figure}

\section{Numerical Results and Discussion}

Each term in Eq.~(\ref{Pis}) has a product of \emph{two} resonance energy denominators that include $x_{\mts i}$ and $x_{\mts s}$, incident and scattered photon energies. The scattering probability increases drastically when both denominators are close to zero for \textit{different} $x_{\mts i}$ and $x_{\mts s}$. This is what makes the Raman scattering cross-section (\ref{cross-sec}) resonant. Additionally, we have the scattering probability amplification factor (\ref{A}) to boost the cross-section amplitude by as much as $\sim\!\xi_y^2(\vc{r}_{A},\omega_p)$ for the TLS strongly coupled to the quasistatic local fields of interband plasmons  (discussed in Ref.~\cite{bondarev2015plasmon}; see also Appendix~C) in the MS near-field zone as well as the angular pre-factor (\ref{AngularFactor}) to control the geometry and polarization of incoming photons relative to the MS plane and SWCN alignment direction.

\begin{figure}[t]
	\begin{centering}
		\includegraphics[width=\textwidth]{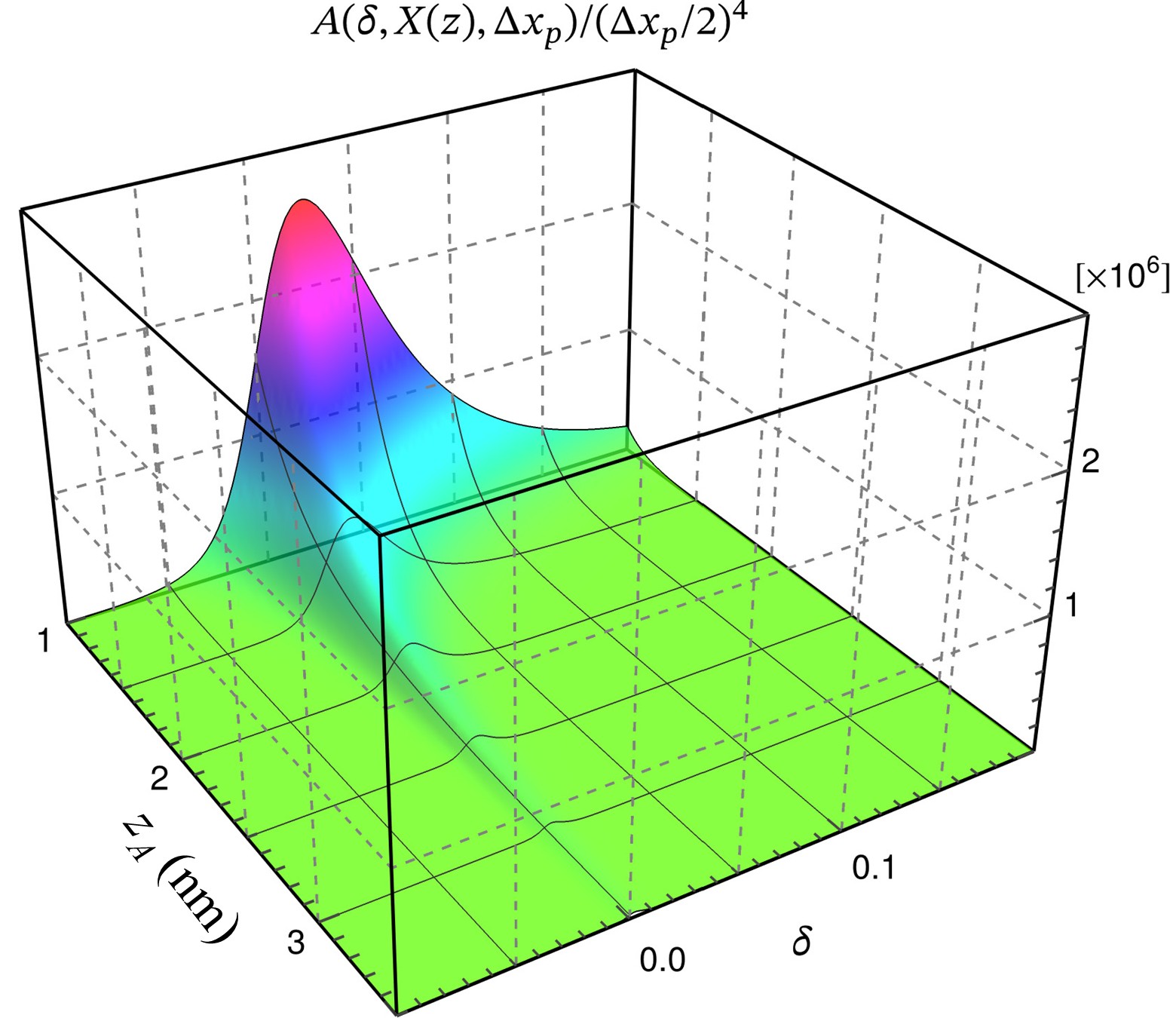}
		\par\end{centering}
	\caption{Amplification factor $A\bigl(\delta,X,\Delta x_{p}\bigr)/(\Delta x_{p}/2)^{4}$ of Eq.~(\ref{A}) with $\Delta x_{p}\!=\!0.004$ showing the influence of detuning $\delta=x_{A}\!-x_{p}$ and distance $z_{A}$ between TLS and MS on the maximum intensity of the plasmon enhanced Raman scattering effect. Here, $\Delta x_{p}$ is obtained from the quantum interband plasmon resonance shown in Fig.~\ref{fig2}, and $z_A$ comes from $X$ due to the distance dependent LDOS calculated as reported in Ref.~\cite{Pugh}.}
	\label{fig5}
\end{figure}

The angular factor $F$ is shown in Fig.~\ref{fig4} (see also Fig.~\ref{fig3} for angle definitions), where it can be seen to reach maximum values for both $p$-polarized ($\eta_{\mts i}\!=\!0$) and $s$-polarized ($\eta_{\mts i}\!=\!\pp/2$) incoming photons at normal incidence ($\theta_{\mts i}\!=0$) when the incidence plane is along ($\phi_{\mts i}\!=\!\pp/2$) or perpendicular ($\phi_{\mts i}\!=\!0$) to the nanotube alignment direction, respectively. Increasing $\theta_{\mts i}$ decreases the angular factor $F$ for $p$-polarized while not changing it for $s$-polarized incoming photons---clearly, because the $p$-polarized photon's electric vector in-plane projection decreases with $\theta_{\mts i}$ while that of the $s$-polarized photon does not change. The effect comes from the anisotropy of the SWCN metasurface where its \textit{virtual} medium-assisted quantum plasma excitations, creating polarizability caused by the incoming photon's electric vector and responsible for the TLS--MS coupling, can only propagate in the SWCN alignment direction. It can also be seen that, while never being zero for $s$-polarized incoming photons, the factor $F$ can be zero for $p$-polarized photons incoming at $\theta_{\mts i}\!=\!\pp/4$ with arbitrarily oriented (any $\phi_{\mts i}$) incidence plane, or for normally incoming photons ($\theta_{\mts i}\!=0$) with $\phi_{\mts i}\!=\eta_{\mts i}\!=\!\pp/4$. In the former case, the in-plane projections of the incoming photon's momentum and electric polarization vector are collinear favoring the surface plasmon excitation, a \textit{real} particle that is created with incoming photon being absorbed. In the latter case, the incoming photon's electric vector turns out being directed perpendicular to the SWCN alignment creating no medium polarizability and no TLS--MS coupling, accordingly. In both cases, there should be no photon scattering for obvious reasons mentioned, which is why the zero angular factor $F$ in Eq.~(\ref{S}) supresses totally the Raman scattering cross-section in Eq.~(\ref{cross-sec}).

\begin{figure}[t]
	\begin{centering}
		\includegraphics[width=\textwidth]{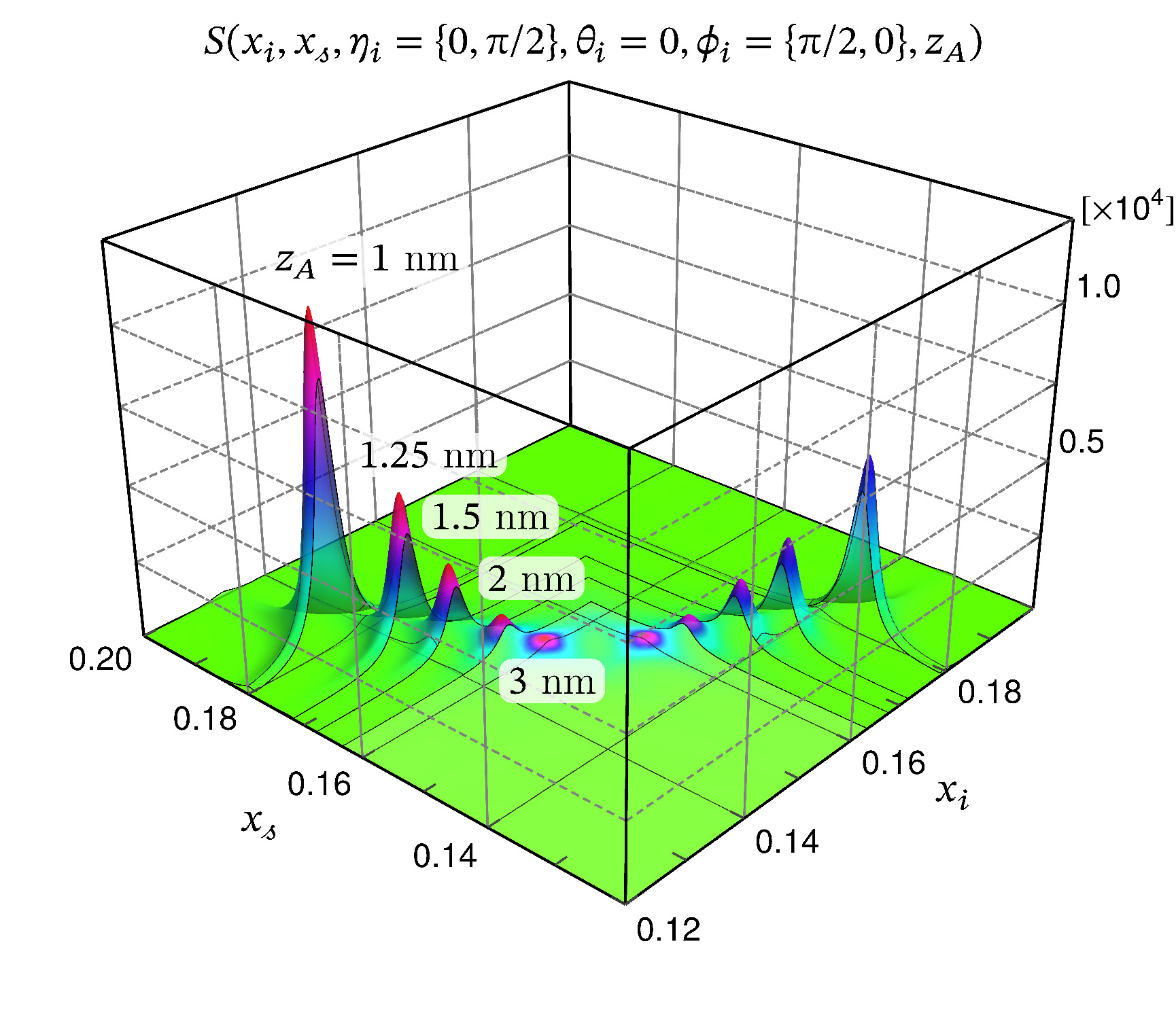}
		\par\end{centering}\vskip-1cm
	\caption{Scattering factor $S$ of the Raman differential cross-section, calculated for normal incidence from Eq.~(\ref{S}) taken at fixed TLS--MS distances $z_A$ as a function of $x_{\mts i}$ and $x_{\mts s}$ for photons of $p$-polarization parallel to the SWCN alignment direction, \emph{or} photons of $s$-polarization perpendicular to it---see Figs.~\ref{fig3} and \ref{fig4} for geometry.}
	\label{fig6}
\end{figure}

For nonzero $F$, the maximal Raman scattering intensity is controlled by the amplification factor $A(\delta,X,\Delta x_p)$ of Eq.~(\ref{A}). This is a function of $\delta$ and $z_A$ shown in Fig.~\ref{fig5} in the form divided by the factor $(\Delta x_p/2)^4$ coming from the resonance denominator in the scattering probability of Eq.~(\ref{Pis}). It was calculated with $\Delta x_{p}\!\approx\!0.004$ obtained from the quantum interband plasmon resonance marked by vertical dotted green line in Fig.~\ref{fig2}. The $z_{A}$ dependence comes from the Rabi splitting parameter $X$ due to the distance dependent LDOS function $\xi_y(\vc{r}_{A},\omega_p)$ we calculated for the same system previously~\cite{Pugh}. We see that the spectral detuning band, originating from the negative refraction band of the EM response in Fig.~\ref{fig2}, can be as large as $\sim\!0.1\times2\gamma_{0}\!=\!0.54$~eV for $z_A\!\lesssim3$~nm to amplify the Raman scattering effect by a factor of up to $10^6$. Clearly, this spectral band can be shifted both to the red and to the blue by varying MS parameters such as SWCN diameters and diameter distribution, offering the flexibility and precise tunability in designing SWCN based SERS substrates.

Returning to Eq.~(\ref{Pis}) we see that for each $x_{\mts i}=x_p+\delta_\pm/2$ only one term contributes, resulting in either Stokes scattering with $x_{\mts s}\!=\!x_p+\delta_-/2<x_{\mts i}\!=\!x_p+\delta_+/2$ or anti-Stokes scattering with $x_{\mts s}\!=\!x_p+\delta_+/2>x_{\mts i}\!=\!x_p+\delta_-/2$. The Raman shift is given by $(\delta_+\!-\delta_-)/2\!=\!\sqrt{\delta^2\!+\!X^2}$, yielding a quantity $\sim\!X$ at resonance, where $\delta\!\sim\!0$ and so $X^2\!/\delta^2\!\gg\!1$, or a quantity $\sim\!\delta$ out of resonance where $X^2\!/\delta^2\!\ll\!1$. Figure~\ref{fig6} shows our $S$-factor numerical calculations of Eq.~(\ref{S}) for a few TLS--MS distances $z_A$ when the angular pre-factor $F$ is maximal ($\phi_{\mts i}\!=\pp/2$ for $p$-polarized and $\phi_{\mts i}\!=0$ for $s$-polarized photons at normal incidence). One can see the Raman scattering effect to exhibit a strong sensitivity to the TLS--MS coupling strength (coming from the Rabi splitting parameter $X$), which decreases with increasing $z_A$, so that the Raman scattering cross-section blows up by a factor of the order of $\sim\!10^4$ for $z_A=1$~nm getting then quenched as $z_A$ increases.

\section{Conclusions}

To summarize, we develop a quantum EM theory for anisotropic SERS phenomena by using a fully-quantized, medium-assisted QED approach for an atom modelled by a TLS in near proximity to an ultrathin closely packed SWCN film. This work extends and further develops the SERS theory previously reported by one of us for TLS near a single-wall carbon nanotube~\cite{bondarev2015plasmon}. Our model provides a unified description of the quantum EM near-field medium-assisted enhancement effects for in-plane anisotropic metasurfaces, of which ultrathin periodically aligned SWCN films are the representative example, and so it gives particular attention to incoming photon parameters of the external light radiation such as photon polarization and incidence plane orientation relative to the main anisotropy axis (CN alignment axis). Our theory applies to atomic type species such as atoms, ions, molecules, or semiconductor quantum dots that are \emph{physisorbed} on the \textit{achiral} SWCN metasurface and strongly coupled to intrinsic EM excitations (interband plasmons) in the near-field zone of the MS structure. The effects of SWCN chirality are not included and will be addressed separately elsewhere. The analytical expression we obtain for the Raman cross-section covers both weak and strong TLS--MS coupling, and shows dramatic SERS enhancement in the strong near-field EM coupling regime. We expect our work to help establish new advanced design concepts for future generation CN based nanophotonics platforms for single molecule/atom/ion detection and optical manipulation in general, to benefit from the extraordinary stability, flexibility and precise tunability of the EM properties of SWCNs by means of their diameter and chirality variation.

\newpage

\section*{Acknowledgments}
I.V.B. is supported by the U.S. Army Research Office grant award No. W911NF2310206.

\section*{Disclosures}
The authors declare no conflicts of interest.

\vspace{\baselineskip}

\begin{appendices}
	\renewcommand{\thesection}{\Alph{section}:}
	\numberwithin{equation}{section}
	\renewcommand{\theequation}{\Alph{section}.\arabic{equation}}
	
	\section{Averaging of transition dipoles}\label{sec:Decomposition-of-Dipole-Factor}
	
	In general, transition dipole moments $d_{\alpha}=\langle u\vert\bwhat d_{\alpha}\vert l\rangle=\langle l\vert\bwhat d_{\alpha}\vert u\rangle^\ast$ can be written in the operator form as follows
	\begin{equation}
		\bwhat d_{\alpha}=d_{\alpha}\bwhat{\sigma}^{\dagger}+d^{\ast}_{\alpha}\bwhat{\sigma},
		\label{dalpha}
	\end{equation}
	where $\alpha$ is a coordinate index and $\bwhat{\sigma}^{\dagger}$, $\bwhat{\sigma}$ are the Pauli operators. Together with the completeness relation
	\begin{align*}
		\bwhat I & =\vert l\rangle\langle l\vert+\vert u\rangle\langle u\vert,
	\end{align*}
	we may then expand and simplify pairs of dipoles as follows
	\begin{align*}
		d_{\lambda}d^{\ast}_{\mu}=d^{\ast}_{\mu}d_{\lambda}=\langle l\vert\bwhat d^{\,\dagger}_{\mu}\vert u\rangle\langle u\vert\bwhat d_{\lambda}\vert l\rangle=\langle l\vert\bwhat d^{\,\dagger}_{\mu}(\bwhat I-\vert l\rangle\langle l\vert)\bwhat d_{\lambda}\vert l\rangle\hskip1.9cm\\
		=\langle l\vert\bwhat d^{\,\dagger}_{\mu}\bwhat d_{\lambda}\vert l\rangle-\langle l\vert\bwhat d^{\,\dagger}_{\mu}\vert l\rangle\langle l\vert(d_{\lambda}\bwhat{\sigma}^{\dagger}+d^{\ast}_{\lambda}\bwhat{\sigma})\vert l\rangle=\langle l\vert\bwhat d^{\,\dagger}_{\mu}\bwhat d_{\lambda}\vert l\rangle\,.
	\end{align*}
	Repeated use of this technique brings the product in Eq.~\eqref{eq:TransitionRate} to
	\begin{equation}
		d_{\lambda}d^{\ast}_{\mu}d_{\nu}d^{\ast}_{\eta}=\langle l\vert\bwhat d^{\,\dagger}_{\mu}\bwhat d_{\lambda}\bwhat d^{\,\dagger}_{\eta}\bwhat d_{\nu}\vert l\rangle\,.
	\end{equation}
	We then substitute
	\begin{equation}
		\bwhat d_{\alpha}=e\bwhat{r}_{\alpha}=e\sqrt{\!\frac{\hbar}{2m\omega_{A}}}\big(\bwhat a_{\alpha}+\bwhat a^{\dagger}_{\alpha}\big)\,,\label{dalphaapm}
	\end{equation}
	where
	\begin{equation*}
		\bwhat a_{\alpha}=\Sqrt{\frac{m\omega_{A}}{2\hbar}}\,\bwhat{r}_{\alpha}+\frac{\ii\bwhat{p}_{\alpha}}{\sqrt{2m\hbar\omega_{A}}}~~~\mbox{and}~~~\bwhat a^{\dagger}_{\alpha} =\Sqrt{\frac{m\omega_{A}}{2\hbar}}\,\bwhat{r}_{\alpha}-\frac{\ii\bwhat{p}_{\alpha}}{\sqrt{2m\hbar\omega_{A}}}
	\end{equation*}
	are canonical ladder operators formed of electron's canonical position and momentum (see, e.g., Ref.~\cite{AbersQM}). This effectively transfers the TLS problem from the two-level atomic space to a Fock space with ground state $\vert l\rangle\!=\!\vert\{0\}\rangle$ representing the zero-excitation state of a quantum atomic oscillator, so that now one can use the commutation
	relations
	\begin{align*}
		[\bwhat a_{\alpha},\bwhat a^{\dagger}_{\beta}] & =\bwhat a_{\alpha}\bwhat a^{\dagger}_{\beta}-\bwhat a^{\dagger}_{\beta}\bwhat a_{\alpha}=\kd{\alpha\beta}
	\end{align*}
	to eventually obtain
	\begin{equation}
		\langle\{0\}\vert(\bwhat a_{\mu}+\bwhat a^{\dagger}_{\mu})(\bwhat a_{\lambda}+\bwhat a^{\dagger}_{\lambda})(\bwhat a_{\eta}+\bwhat a^{\dagger}_{\eta})(\bwhat a_{\nu}+\bwhat a^{\dagger}_{\nu})\vert\{0\}\rangle=\kd{\mu\eta}\kd{\lambda\nu}+\kd{\lambda\eta}\kd{\mu\nu}+\kd{\mu\lambda}\kd{\eta\nu}\,,\label{dddd}
	\end{equation}
	yielding
	\begin{equation*}
		d_{\lambda}d^{\ast}_{\mu}d_{\nu}d^{\ast}_{\eta}=\biggl(\frac{e^2\hbar}{2m\omega_{A}}\biggl)^2\Big(\kd{\mu\eta}\kd{\lambda\nu}+\kd{\lambda\eta}\kd{\mu\nu}+\kd{\mu\lambda}\kd{\eta\nu}\Big)\,.
	\end{equation*}
	Note also that
	\begin{equation}
		d_{\lambda}d^{\ast}_{\mu}=\langle l\vert\bwhat d^{\,\dagger}_{\mu}\bwhat d_{\lambda}\vert l\rangle =\frac{e^{2}\hbar}{2m\omega_{A}}\kd{\mu\lambda}\label{dd}
	\end{equation}
	by the above technique. It then follows directly that
	\begin{equation}
		|d|^{2}=\sum_{\lambda}d_{\lambda}d^{\ast}_{\lambda}=\frac{3e^{2}\hbar}{2m\omega_{A}}\,,\label{dabs2}
	\end{equation}
	so that Eq.~(\ref{dddd}) takes the form
	\begin{equation}
		d_{\lambda}d^{\ast}_{\mu}d_{\nu}d^{\ast}_{\eta}=\frac{|d|^4}{3^2}\Big(\kd{\mu\eta}\kd{\lambda\nu}+\kd{\lambda\eta}\kd{\mu\nu}+\kd{\mu\lambda}\kd{\eta\nu}\Big)\,.
	\end{equation}
	With this expression, one obtains
	\begin{equation}
		\biggl|\bsum_{\lambda,\mu,\nu,\eta}\!\!\!d_{\lambda}d^{\ast}_{\mu}d_{\nu}d^{\ast}_{\eta}\uscr s{\lambda}^{\ast}\uscr i{\mu}\mts G_{\nu\eta}\,\biggr|^{2}\!=\frac{|d|^8}{3^4}\,\biggl|\,\bsum_{\lambda,\mu}\Big\{\uscr s{\lambda}^{\ast}\uscr i{\mu}\big[\mts G_{\lambda\mu}+\mts G_{\mu\lambda}\big]+\uscr s{\lambda}^{\ast}\uscr i{\lambda}\mts G_{\mu\mu}\Big\}\,\biggr|^{2},
	\end{equation}
	which by introducing the traceless symmetric tensor
	\begin{equation*}
		\overline{\mts G}_{\lambda\mu}=\frac{1}{2}\biggl(\mts G_{\lambda\mu}+\mts G_{\mu\lambda}-\frac{2}{3}\kd{\lambda\mu}\tr\mts G\biggr)\,,
	\end{equation*}
	in it, transforms to Eq.~(\ref{ddddfinal}).
	
	\vspace{\baselineskip}
	
	\section{Angular Factor Derivation}\label{sec:Polarization-Derivation}
	
	We refer to the schematic presented in Fig.~\ref{fig3}, which also specifies the notations we use. An incident photon of unit polarization vector $\uvcr i$ (complex in general) with wave vector $\vc k_{\mts i}=k_{\mts i}\hat{\vc k}_{\mts i}$ making angle of incidence $\theta_{\mts i}$ with the $z$-axis, is scattered by the coupled TLS--MS system residing in the $xy$-plane $\mathcal F$ with CN axes aligned in the $y$-direction. The scattered photon of unit polarization vector $\uvcr s$ has wave vector $\vc k_{\mts s}=k_{\mts s}\hat{\vc k}_{\mts s}$. The plane of incidence $\mathcal P$ is spanned by $\hat{\vc k}_{\mts i}$ and the $z$-axis and makes the angle $\phi_{\mts i}$ with the $x$-axis. In spherical coordinates,
	\begin{align*}
		\hat{\vc k}_{\mts i} & =(\sin\tilde{\theta}_{\mts i}\cos\phi_{\mts i},\sin\tilde{\theta}_{\mts i}\sin\phi_{\mts i},\cos\tilde{\theta}_{\mts i})
	\end{align*}
	but substituting $\tilde{\theta}_{\mts i}=\pp-\theta_{\mts i}$ makes this
	\begin{align*}
		\hat{\vc k}_{\mts i} & =(\sin\theta_{\mts i}\cos\phi_{\mts i},\sin\theta_{\mts i}\sin\phi_{\mts i},-\cos\theta_{\mts i})\,.
	\end{align*}
	The unit vector $\hat{\vc s}$, which is orthogonal to the $\mathcal P$-plane, can then be found as:
	\begin{align*}
		\hat{\vc s} & =\frac{\hat{\vc k}_{\mts i}\times\hat{\vc z}}{|\hat{\vc k}_{\mts i}\times\hat{\vc z}|}=(\sin\phi_{\mts i},-\cos\phi_{\mts i},0).
	\end{align*}
	The unit vector $\hat{\vc p}_{\mts i}$, lying in the $\mathcal P$-plane, can be constructed similarly:
	\begin{align*}
		\hat{\vc p}_{\mts i} & =\hat{\vc s}\times\hat{\vc k}_{\mts i}=(\cos\theta_{\mts i}\cos\phi_{\mts i},\cos\theta_{\mts i}\sin\phi_{\mts i},\sin\theta_{\mts i}).
	\end{align*}
	Now in possession of an orthonormal basis vector set $(\hat{\vc k}_{\mts i},\hat{\vc p}_{\mts i},\hat{\vc s})$,
	we can write the polarization vector of the incident photon as
	\begin{align*}
		\uvcr i & =\cos\eta_{\mts i}\hat{\vc p}_{\mts i}+\sin\eta_{\mts i}\hat{\vc s}
	\end{align*}
	with $\eta_{\mts i}=\pp/2$ and $\eta_{\mts i}=0$ indicating the $s$- and $p$-polarizations, where an electric field is parallel
	to $\hat{\vc s}$ and $\hat{\vc p}_{\mts i}$, respectively, of transverse electric (TE) and transverse magnetic (TM) plane waves.
	
	For incoming photons with wavelengths exceeding the translational period of an SWCN film, it is natural to assume that $\hat{\vc k}_{\mts s}$ also lies in the $\mathcal P$-plane so that $\phi_{\mts s\!}=\phi_{\mts i}$. Moreover,
	neglecting the Goos-H\"{a}nchen and Fedorov-Imbert effects out of scope here~\cite{BiehsBond2025,BliokhAiello13}, the scattered angle $\theta_{\mts s\!}=\theta_{\mts i}$ so that
	\begin{align*}
		\hat{\vc k}_{\mts s} & =(\sin\theta_{\mts i}\cos\phi_{\mts i},\sin\theta_{\mts i}\sin\phi_{\mts i},\cos\theta_{\mts i}),
	\end{align*}
	whereby
	\begin{align*}
		\hat{\vc p}_{\mts s} & =\hat{\vc s}\times\hat{\vc k}_{\mts s}=(-\cos\theta_{\mts i}\cos\phi_{\mts i},-\cos\theta_{\mts i}\sin\phi_{\mts i},\sin\theta_{\mts i})\,.
	\end{align*}
	Finally, for \textit{achiral} SWCN films considered here, we also have $\eta_{\mts s}=\eta_{\mts i}$, hence \begin{align*}
		\uvcr s & =\cos\eta_{\mts i}\hat{\vc p}_{\mts s}+\sin\eta_{\mts i}\hat{\vc s}.
	\end{align*}
	
	One can now, after simplifications, find the angular factor
	\begin{align*}
		F(\eta_{\mts i},\phi_{\mts i},\theta_{\mts i})=\lhs{\left|\uscr sx^{\ast}\uscr ix+3\uscr sy^{\ast}\uscr iy+\uscr sz^{\ast}\uscr iz\right|^{2}}
	\end{align*}
	in the form presented in Eq.~(\ref{AngularFactor}).
	
	\vspace{\baselineskip}
	
	\section{Energy Levels and Mixing Coefficients}\label{sec:Formulation-of-Mixing-Coefficients}
	
	Inserting the dipole average~(\ref{dd}) in Eq.~(\ref{Eint}), one obtains
	\begin{eqnarray}
		E\!&=&\!\frac{\hbar\omega_{A}}{2}+\frac{4\hbar|d|^2}{3c^{2}}\!\!\Int 0{\infty}\!\!\d{\omega}\omega^{2}\frac{\sum_{\mu}\Im G_{\mu\mu}(\vc r_{A},\vc r_{A},\omega)}{\hbar\omega_{A}/2-\hbar\omega+E}\nonumber\\
		&\approx&\!\frac{\hbar\omega_{A}}{2}+\frac{4\hbar|d|^2}{3c^{2}}\!\!\Int 0{\infty}\!\!\d{\omega}\omega^{2}\frac{\Im G^{\mathrm{sc}}_{yy}(\vc r_{A},\vc r_{A},\omega)}{\hbar\omega_{A}/2-\hbar\omega+E}
		\label{Eapprox}
	\end{eqnarray}
	due to the fact that $\sum_{\mu}\Im G_{\mu\mu}(\vc r_{A},\vc r_{A},\omega)\!\approx\!\Im G^{\mathrm{sc}}_{yy}(\vc r_{A},\vc r_{A},\omega)$ under the resonance conditions, where $\vc r_{A}$ is in the evanescent field zone and $\hbar\omega$ is in the vicinity of a quantum interband plasmon resonance $\hbar\omega_p$ of the SWCN metasurface (marked by the dotted vertical green line in Fig.~\ref{fig2}). Details can be found in Ref.~\cite{Pugh}, where the near-surface local photonic density-of-states (LDOS) function $\xi_y(\vc r_{A},\omega)\!=\!2\pp c\Im G^{\mathrm{sc}}_{yy}(\vc r_{A},\vc r_{A},\omega)/(\omega\sqrt{\varepsilon_2})$ associated with scattering part $\Im G^{\mathrm{sc}}_{yy}(\vc r_{A},\vc r_{A},\omega)$ of the $\Im G_{yy}$ tensor component in the TLS location region of dielectric permittivity $\varepsilon_2$ (=1 for air), is shown to have a highly resonant behavior under these conditions. We then apply a Lorentzian approximation in the integral above by setting up
	\[
	\Im G^{\mathrm{sc}}_{yy}(\vc r_{A},\vc r_{A},\omega)=\frac{\omega\sqrt{\varepsilon_2}}{2\pp c}\xi_y(\vc r_{A},\omega)\approx\frac{3\hbar c^2\Gamma_0(\omega)}{8\pp|d|^2\omega^2}\frac{\xi_y(\vc r_{A},\omega_p)\Delta\omega_0^2}{(\omega-\omega_p)^2+\Delta\omega_0^2}
	\]
	with $\Delta\omega_0$ being the half-width-at-half-maximum of the Lorentz-line approximated LDOS and
	\begin{equation}
		\Gamma_0(\omega)=\frac{4|d|^2\omega^3\sqrt{\varepsilon_2}}{3\hbar c^3}\label{G0}
	\end{equation}
	representing the isotropic dipolar spontaneous emission rate in free dielectric space, whereby the second term in Eq.~(\ref{Eapprox}) takes the form
	\[
	\frac{4\hbar|d|^2}{3c^{2}}\!\!\Int 0{\infty}\!\!\d{\omega}\omega^{2}\frac{\Im G^{\mathrm{sc}}_{yy}(\vc r_{A},\vc r_{A},\omega)}{\hbar\omega_{A}/2-\hbar\omega+E}\approx\frac{\hbar^2}{2\pp}\frac{\Gamma_0(\omega)\xi_y(\vc r_{A},\omega_p)\Delta\omega_0^2}{\hbar\omega_{A}/2-\hbar\omega_p+E}\!\Int 0{\infty}\!\!\frac{\d{\omega}}{(\omega-\omega_p)^2+\Delta\omega_0^2}\,.
	\]
	Here, the integral calculates to give $[\arctan(\omega_p/\Delta\omega_{0})+\pp/2]/\Delta\omega_{0}$, yielding $\pp/\Delta\omega_{0}$ with the $\arctan$-function expanded to linear terms in $\Delta\omega_{0}/\omega_p$ ($\ll\!1$, and the stronger this inequality is, the better such a series expansion works). Equation~(\ref{Eapprox}) now becomes a quadratic equation yielding
	\begin{equation}
		\varepsilon_{1,2}=\frac{1}{2}\left(x_p\mp\sqrt{\delta^2+X^2}-i\Delta x_p\right)
		\label{2en}
	\end{equation}
	Here, $\varepsilon_{1,2}=E_{1,2}/(2\gamma_{0})$ and $(x_A,x_p,\Delta x_p)=\hbar(\omega_A,\omega_p,\Delta\omega_p)/2\gamma_{0}$ are dimensionless energies, where $\gamma_0=2.7$~eV is the carbon-carbon overlap integral of graphene,
	\begin{equation}
		\delta\!=\!x_A-x_p\;\;\;\mbox{and} \;\;\;X\!=\!\frac{\hbar}{2\gamma_{0}}\sqrt{2\Delta\omega_{0}\Gamma_0(\omega_p)\xi_y(\vc{r}_{A},\omega_p)}
		\label{X}
	\end{equation}
	are responsible for the TLS--MS coupling strength, and~$\Delta x_p$ is added to phenomenologically account for the finite half-width of the plasmon resonance (finite plasmon lifetime), as it can be seen in Fig.~\ref{fig2}, which is assumed to be much broader than the excited atomic level natural half-width dropped here on this account for simplicity.
	
	By the same technique, plugging Eq.~(\ref{2en}) in Eq.~(\ref{Cu}), after simplifications we obtain
	\begin{align*}
		|C^{(1,2)}_{u}|^{2}=\frac{1}{2}\left(1+\frac{1\mp\sqrt{1+X^{2}/\delta^{2}}}{1+X^{2}/\delta^{2}\mp\sqrt{1+X^{2}/\delta^{2}}}\right),
	\end{align*}
	whence
	\begin{equation}
		C^{(1,2)}_{u}=\left[\frac{1}{2}\left(1+\frac{1\mp\sqrt{1+X^{2}/\delta^{2}}}{1+X^{2}/\delta^{2}\mp\sqrt{1+X^{2}/\delta^{2}}}\right)\right]^{1/2}\label{Cu12approx}
	\end{equation}
	and
	\begin{equation}
		|C^{(2)}_{u}|^{2}|C^{(1)}_{u}|^{2}=|C^{(2)}_{u}C^{(1)}_{u}|^{2}=\frac{X^2/4}{\delta^2+X^2},\label{Cu2Cu1}
	\end{equation}
	accordingly. Additionally, from the normalization condition (\ref{norm}), we now have
	\begin{equation}
		\Int 0{\infty}\!\d{\omega}\!\int\!\d{\vc{\rho}}\sum_{\lambda}|C^{(1,2)}_{l\lambda}(\vc{\rho},\omega)|^{2}=1-|C_u^{(1,2)}|^2\approx\frac{(X^2/2\delta^2)|C_u^{(1,2)}|^2}{1+X^2/2\delta^2\mp\sqrt{1+X^2/\delta^2}}\,.\label{C12sqd}
	\end{equation}
	
	Equations (\ref{2en})--(\ref{C12sqd}) are valid both in resonance, where $\delta\!\sim\!0$ and so $X^2/\delta^2\!\gg\!1$, and out of resonance where $X^2/\delta^2\!\ll\!1$, and provide different asymptotic expressions in these two regimes of relevance to strong and weak TLS--MS coupling, respectively. As was previously discussed by one of us for TLS coupled to an individual SWCN~\cite{bondarev2015plasmon}, the actual coupling regime, strong or weak, depends on the relation between $X$ and $\Delta x _p$. When $X^2/\delta^2\!\gg\!1$, the TLS--MS system can only be coupled strongly if $X$, which is the medium-assisted vacuum Rabi splitting as one can see from Eq.~(\ref{2en}), is much greater than the plasmon broadening $\Delta x_p$. In this case, the Rabi splitting of the $\varepsilon_1$ and $\varepsilon_2$ levels is not hidden by the plasmon broadening. Otherwise, these levels are smeared, show no clear anti-crossing behavior and no strong coupling, accordingly.
	
\end{appendices}


\begin{thebibliography}{99}
	
	\bibitem{stiles2008surface}P. L. Stiles, J. A. Dieringer, N. C. Shah, and R. P. Van Duyne, Surface-enhanced raman spectroscopy, {Annu. Rev. Anal. Chem.} \textbf{1}, 601 (2008).
	
	\bibitem{acsnano}J. Langer, D. Jimenez de Aberasturi, J. Aizpurua, \emph{et al.}, Present and future of surface-enhanced Raman scattering, {ACS Nano} \textbf{14}, 28 (2020).
	
	\bibitem{han2021surface}X. X. Han, R. S. Rodriguez, C. L. Haynes, \emph{et al.}, Surface-enhanced Raman spectroscopy, {Nat. Rev. Methods Primers} \textbf{1}, 87 (2022).
	
	\bibitem{campion1998surface}A. Campion and P. Kambhampati, Surface-enhanced Raman scattering, {Chem. Soc. Rev.} \textbf{27}, 241 (1998).
	
	\bibitem{mccreery2005raman}R. L. McCreery, \emph{Raman spectroscopy for chemical analysis} (John Wiley \& Sons, Ltd, 2000).
	
	\bibitem{abdelsalam2005electrochemical}M. E. Abdelsalam, P. N. Bartlett, J. J. Baumberg, \emph{et al.}, Electrochemical SERS at a structured gold surface,{Electrochem. Commun.} \textbf{7}, 740 (2005).
	
	\bibitem{hartman2016surface}T. Hartman, C. S. Wondergem, N. Kumar, \emph{et al.}, Surface- and tip-enhanced Raman spectroscopy in catalysis, {J. Phys. Chem. Lett.} \textbf{7}, 1570 (2016).
	
	\bibitem{medical_app}W. Xie and S. Schlücker, Medical applications of surface-enhanced Raman scattering, {Phys. Chem. Chem. Phys.} \textbf{15}, 5329 (2013).
	
	\bibitem{jamieson2017bioanalytical}L. E. Jamieson, S. M. Asiala, K. Gracie, \emph{et al.}, Bioanalytical measurements enabled by surface-enhanced Raman scattering (SERS) probes,{Annu. Rev. Anal. Chem.} \textbf{10}, 415 (2017).
	
	\bibitem{SHARMA201216}B. Sharma, R. R. Frontiera, A.-I. Henry, \emph{et al.}, SERS: Materials, applications, and the future, {Mater. Today} \textbf{15}, 16 (2012).
	
	\bibitem{mcquillan2009discovery}A. J. McQuillan, Recollection. The discovery of surface-enhanced Raman scattering, {Notes and Rec.} \textbf{63}, 209 (2009).
	
	\bibitem{jeanmaire1977surface}D. L. Jeanmaire and R. P. {Van Duyne}, Surface Raman spectroelectrochemistry: Part i. Heterocyclic, aromatic, and aliphatic amines adsorbed on the anodized silver electrode, {J. Electroanal. Chem. Interfacial Electrochem.} \textbf{84}, 1 (1977).
	
	\bibitem{lee2007surface}S. J. Lee, Z. Guan, H. Xu, and M. Moskovits, Surface-enhanced Raman spectroscopy and nanogeometry: The plasmonic origin of sers, {J. Phys. Chem. C} \textbf{111}, 17985 (2007).
	
	\bibitem{ding2017electromagnetic}S.-Y. Ding, E.-M. You, Z.-Q. Tian, and M. Moskovits, Electromagnetic theories of surface-enhanced raman spectroscopy, {Chem. Soc. Rev.} \textbf{46}, 4042 (2017).
	
	\bibitem{schedin2010surface}F. Schedin, E. Lidorikis, A. Lombardo, \emph{et al.}, Surface-enhanced Raman spectroscopy of graphene, {ACS Nano} \textbf{4}, 5617 (2010).
	
	\bibitem{xu2012surface}W. Xu, X. Ling, J. Xiao, \emph{et al.}, Surface enhanced Raman spectroscopy on a flat graphene surface, {Proc. Natl. Acad. Sci.} \textbf{109}, 9281 (2012).
	
	\bibitem{lai2018recent}H. Lai, F. Xu, Y. Zhang, and L. Wang, Recent progress on graphene-based substrates for surface-enhanced Raman scattering applications, {J. Mater. Chem. B} \textbf{6}, 4008 (2018).
	
	\bibitem{Saito}R. Saito, G. Dresselhaus, and M. S. Dresselhaus, \emph{Science of Fullerens and Carbon Nanotubes} (Imperial College, 1998).
	
	\bibitem{bondarev2015plasmon}I. V. Bondarev, Plasmon enhanced Raman scattering effect for an atom near a carbon nanotube, {Opt. Express} \textbf{23}, 3971 (2015).
	
	\bibitem{Pendry98}F. J. Garsía-Vidal, J. M. Pitarke, and J. B. Pendry, Silver-filled nanotubes used as spectroscopic enhancers, Phys. Rev. B 58, 6783 (1998).
	
	\bibitem{Optical_Mater_1576}J. Zaumseil, Luminescent defects in single-walled carbon nanotubes for applications, {Adv. Opt. Mater.} \textbf{10}, 2101576 (2022).
	
	\bibitem{Natute_commu4439}Y. Zheng, Y. Han, B. M. Weight, \emph{et al.}, Photochemical spin-state control of binding configuration for tailoring organic color center emission in carbon nanotubes, {Nat. Commun.} \textbf{13}, 4439 (2022).
	
	\bibitem{ACS_Nano_11742}M.-K. Li, A. Riaz, M. Wederhake, \emph{et al.}, Electroluminescence from single-walled carbon nanotubes with quantum defects, {ACS Nano} \textbf{16}, 11742 (2022).
	
	\bibitem{Nature_chem1089}A. Saha, B. J. Gifford, X. He, \emph{et al.}, Narrow-band single-photon emission through selective aryl functionalization of zigzag carbon nanotubes, {Nat. Chem.} \textbf{10}, 1089 (2018).
	
	\bibitem{Nature_photonics727}S. Khasminskaya, F. Pyatkov, K. S{\l}owik, \emph{et al.}, Fully integrated quantum photonic circuit with an electrically driven light source, {Nat. Photonics} \textbf{10}, 727 (2016).
	
	\bibitem{Nature_Nanotech671}X. Ma, N. F. Hartmann, J. K. S. Baldwin, \emph{et al.}, Room-temperature single-photon generation from solitary dopants of carbon nanotubes, {Nat. Nanotechnol.} \textbf{10}, 671 (2015).
	
	\bibitem{Science850}L. Liu, J. Han, L. Xu, \emph{et al.}, Aligned, high-density semiconducting carbon nanotube arrays for high-performance electronics, {Science} \textbf{368}, 850 (2020).
	
	\bibitem{ScienceAdavance_1240}G. J. Brady, A. J. Way, N. S. Safron, \emph{et al.}, Quasi-ballistic carbon nanotube array transistors with current density exceeding si and gaas, {Sci. Adv.} \textbf{2}, e1601240 (2016).
	
	\bibitem{Functional_material5018}H.-N. Fan, S.-L. Chen, X.-H. Chen, \emph{et al.}, 3d selenium sulfide@carbon nanotube array as long-life and high-rate cathode material for lithium storage, {Adv. Funct. Mater.} \textbf{28}, 1805018 (2018).
	
	\bibitem{Nano_Letter5832}J. A. Roberts, P.-H. Ho, S.-J. Yu, and J. A. Fan, Electrically driven hyperbolic nanophotonic resonators as high speed, spectrally selective thermal radiators, {Nano Letters} \textbf{22}, 5832--5840 (2022).
	
	\bibitem{Adv_Energy1814}Z. Zhang, L. Wang, Y. Li, \emph{et al.}, Nitrogen-doped core-sheath carbon nanotube array for highly stretchable supercapacitor, {Adv. Energy Mater.} \textbf{7}, 1601814 (2017).
	
	\bibitem{small1150}L. Qiu, Q. Wu, Z. Yang, \emph{et al.}, Free-standing aligned carbon nanotube array grown on a large-area single-layered graphene sheet for efficient dye-sensitized solar cell, {Small} \textbf{11}, 1150 (2015).
	
	\bibitem{ChemP132}T. Hertel and I. Bondarev, Photophysics of carbon nanotubes and nanotube composites, {Chem. Phys.} \textbf{413}, 1 (2013).
	
	\bibitem{Nature633}X. He, W. Gao, L. Xie, \emph{et al.}, Wafer-scale monodomain films of spontaneously aligned single-walled carbon nanotubes, {Nat. Nanotechnol.} \textbf{11}, 633 (2016).
	
	\bibitem{Kono2019}W. Gao, C. F. Doiron, X. Li, J. Kono, and G. V. Naik, Macroscopically aligned carbon nanotubes as a refractory platform for hyperbolic thermal emitters, ACS Photonics 6, 1602 (2019)
	
	\bibitem{NanoLett641}K.-C. Chiu, A. L. Falk, P.-H. Ho, \emph{et al.}, Strong and broadly tunable plasmon resonances in thick films of aligned carbon nanotubes, {Nano Letters} \textbf{17}, 5641 (2017).
	
	\bibitem{PNAS115}P.-H. Ho, D. B. Farmer, G. S. Tulevski, \emph{et al.}, Intrinsically ultrastrong plasmon–exciton interactions in crystallized films of carbon nanotubes, {Proc. Natl. Acad. Sci.} \textbf{115}, 12662 (2018).
	
	\bibitem{NanoLett19}M. E. Green, D. A. Bas, H.-Y. Yao, \emph{et al.}, Bright and ultrafast photoelectron emission from aligned single-wall carbon nanotubes through multiphoton exciton resonance, {Nano Letters} \textbf{19}, 158 (2019).
	
	\bibitem{ACS1602}W. Gao, C. F. Doiron, X. Li, \emph{et al.}, Macroscopically aligned carbon nanotubes as a refractory platform for hyperbolic thermal emitters, {ACS Photonics} \textbf{6}, 1602 (2019).
	
	\bibitem{NanoLett3131}J. A. Roberts, S.-J. Yu, P.-H. Ho, \emph{et al.}, Tunable hyperbolic metamaterials based on self-assembled carbon nanotubes, {Nano Letters} \textbf{19}, 3131 (2019).
	
	\bibitem{Vac015}S. Schöche, P.-H. Ho, J. A. Roberts, \emph{et al.}, Mid-IR and UV-vis-NIR M\"{u}ller matrix ellipsometry characterization of tunable hyperbolic metamaterials based on self-assembled carbon nanotubes, {J. Vac. Sci. Technol. B} \textbf{38}, 014015 (2020).
	
	\bibitem{NanoLet20}N. Komatsu, M. Nakamura, S. Ghosh, \emph{et al.}, Groove-assisted global spontaneous alignment of carbon nanotubes in vacuum filtration, {Nano Letters} \textbf{20}, 2332 (2020).
	
	\bibitem{PhyRevApp006}J. A. Roberts, P.-H. Ho, S.-J. Yu, \emph{et al.}, Multiple tunable hyperbolic resonances in broadband infrared carbon-nanotube metamaterials, {Phys. Rev. Appl.} \textbf{14}, 044006 (2020).
	
	\bibitem{WGao2023}J. Doumani, M. Lou, O. Dewey, \emph{et al.}, Engineering chirality at wafer scale with ordered carbon nanotube architectures, {Nature Commun.} \textbf{14}, 7380 (2023).
	
	\bibitem{DipJariwala}J. Lynch, et al., Electrically tunable excitonic-hyperbolicity in chirality-pure carbon nanotubes, E-print arXiv: 2509.24848 (29 Sep 2025).
	
	\bibitem{BondPRappl2021}I. V. Bondarev and C. M. Adhikari, Collective excitations and optical response of ultrathin carbon-nanotube films, {Phys. Rev. Appl.} \textbf{15}, 034001 (2021).
	
	\bibitem{AdhBondJAL2021}C. Adhikari and I. V. Bondarev, Controlled exciton-plasmon coupling in a mixture of ultrathin periodically aligned single-wall carbon nanotube arrays, {J. Appl. Phys.,} \textbf{129}, 015301 (2021).
	
	\bibitem{BondCPC2023}I. V. Bondarev, et al., Confinement-induced nonlocality and Casimir force in transdimensional systems, {Phys. Chem. Chem. Phys.} \textbf{25}, 29257 (2023).
	
	\bibitem{Pablo}P. Rodriguez-Lopez, D.-N. Le, I. V. Bondarev, et al., Giant anisotropy and Casimir phenomena: The case of carbon nanotube metasurfaces, {Phys. Rev. B,} \textbf{109}, 035422 (2024).
	
	\bibitem{Pugh}M. D. Pugh, S. F. Islam, and I. V. Bondarev, Anisotropic photon emission enhancement near carbon nanotube metasurfaces, {Physical Review B} \textbf{109}, 235430 (2024).
	
	\bibitem{ACS_Nano7771}Z. M. Abd El-Fattah, V. Mkhitaryan, J. Brede, \emph{et al.}, Plasmonics in atomically thin crystalline silver films, {ACS Nano} \textbf{13}, 7771 (2019).
	
	\bibitem{Nature_photon899}F. Xia, H. Wang, D. Xiao, \emph{et al.}, Two-dimensional material nanophotonics, {Nat. Photonics} \textbf{8}, 899 (2014).
	
	\bibitem{Nature_photon216}K. F. Mak and J. Shan, Photonics and optoelectronics of 2d semiconductor transition metal dichalcogenides, {Nat. Photonics} \textbf{10}, 216 (2016).
	
	\bibitem{Nat_Commu1762}A. M. Dubrovkin, B. Qiang, H. N. S. Krishnamoorthy, \emph{et al.}, Ultra-confined surface phonon polaritons in molecular layers of van der waals dielectrics, {Nat. Commun.} \textbf{9}, 1762 (2018).
	
	\bibitem{Nano_Lett984}E. L. Runnerstrom, K. P. Kelley, T. G. Folland, \emph{et al.}, Polaritonic hybrid-epsilon-near-zero modes: Beating the plasmonic confinement vs propagation-length trade-off with doped cadmium oxide bilayers, {Nano Letters} \textbf{19}, 948 (2019).
	
	\bibitem{ACS_Photon2816}D. Shah, A. Catellani, H. Reddy, \emph{et al.}, Controlling the plasmonic properties of ultrathin TiN films at the atomic level, {ACS Photonics} \textbf{5}, 2816 (2018).
	
	\bibitem{PRB121408}S. Campione, I. Brener, and F. Marquier, Theory of epsilon-near-zero modes in ultrathin films, {Phys. Rev. B} \textbf{91}, 121408 (2015).
	
	\bibitem{BondMousShal2020}I. V. Bondarev, H. Mousavi, and V. M. Shalaev, Transdimensional epsilon-near-zero modes in planar plasmonic nanostructures, {Phys. Rev. Research} \textbf{2}, 013070 (2020).
	
	\bibitem{BondShal2017}I. V. Bondarev and V. M. Shalaev, Universal features of the optical properties of ultrathin plasmonic films, {Optical Mater. Express} \textbf{7}, 3731 (2017).
	
	\bibitem{MRSC2018}I. V. Bondarev, H. Mousavi, and V. M. Shalaev, Optical response of finite-thickness ultrathin plasmonic films, {MRS Commun.,} \textbf{8}, 1092 (2018).
	
	\bibitem{Bond2019OMEX}I. V. Bondarev, Finite-thickness effects in plasmonic films with periodic cylindrical anisotropy [Invited], {Optical Mater. Express} \textbf{9}, 285 (2019).
	
	\bibitem{Keldysh}L. V. Keldysh, Coulomb interaction in thin semiconductor and semimetal films, {Pisma Zh. Eksp. Teor. Fiz.} \textbf{29}, 716 (1979) [Engl. translation: JETP Lett. \textbf{29}, 658 (1979)].
	
	\bibitem{Rytova}N. S. Rytova, Screened potential of a point charge in a thin film, {Moscow Univ. Phys. Bull.} \textbf{3}, 30 (1967).
	
	\bibitem{BoltShal2019}A. Boltasseva and V. M. Shalaev, Transdimensional photonics, {ACS Photon.} \textbf{6}, 1 (2019).
	
	\bibitem{Shah2022}D. Shah, et al., Thickness-dependent Drude plasma frequency in transdimensional plasmonic TiN, {Nano Lett.} \textbf{22}, 4622 (2022).
	
	\bibitem{Rivera}N. Rivera, et al., Shrinking light to allow forbidden transitions on the atomic scale, {Science} \textbf{353}, 263 (2016).
	
	\bibitem{BondAnnPhys2023}I. V. Bondarev, Controlling single-photon emission with ultrathin transdimensional plasmonic films, {Ann. Phys. (Berlin)} \textbf{535}, 2200331 (2023).
	
	\bibitem{NonlocRoadmap}I. V. Bondarev, et al., Nonlocal effects in transdimensional plasmonics, In: Nonlocality in Photonic Materials and Metamaterials: Roadmap, {Optical Mater. Express} \textbf{15}, 1544 (2025).
	
	\bibitem{BoltRoadmap2025}A. Boltasseva, et al., Transdimensional materials as a new platform for strongly correlated systems, In: Roadmap for photonics with 2D materials, {ACS Photon.} \textbf{12}, 3961 (2025).
	
	\bibitem{Zundel}L. Zundel, et al., Comparative analysis of the near- and far-field optical response of thin plasmonic nanostructures, {Adv. Optical Mater.} \textbf{10}, 2102550 (2022).
	
	\bibitem{BiehsBond2023}S.-A. Biehs and I. V. Bondarev, Far- and near-field heat transfer in transdimensional plasmonic film systems, {Adv. Optical Mater.} \textbf{11}, 2202712 (2023).
	
	\bibitem{Salihoglu2023}H. Salihoglu, et al., Nonlocal near-field radiative heat transfer by transdimensional plasmonics, {Phys. Rev. Lett.} \textbf{131}, 086901 (2023).
	
	\bibitem{Das2024}P. Das, et al., Electron confinement-induced plasmonic breakdown in metals, {Science Adv.} \textbf{10}, eadr2596 (2024).
	
	\bibitem{Crystalliz2026}I. V. Bondarev, A. Boltasseva, J. B. Khurgin, and V. M. Shalaev, Crystallization of the transdimensional electron liquid, {Nano Lett.,} \textbf{26}, 3649 (2026).
	
	\bibitem{BiehsBond2025}S.-A. Biehs and I. V. Bondarev, Goos-H\"{a}nchen effect singularities in transdimensional plasmonic films, {Nanophotonics} \textbf{14}, 4513 (2025).
	
	\bibitem{Taiwan}F.-T. Tseng, et al., Confinement-induced nonlocality and optical non-linearity of transdimensional titanium nitride in the epsilon-near-zero region, E-print arXiv: 2512.14035 (17 Dec 2025).
	
	\bibitem{Science2013}A. V. Kildishev, A. Boltasseva, and V. M. Shalaev, Planar photonics with metasurfaces, {Science} \textbf{339}, 1289 (2013).
	
	\bibitem{ChinaTutorial}Z. Guo, H. Jiang, and H. Chen, Hyperbolic metamaterials: From dispersion manipulation to applications, {J. Appl. Phys.} \textbf{127}, 071101 (2020).
	
	\bibitem{Vasya2025}Z. Liu, T. Nishihara, V. Perebeinos, and Y. Miyauchi, Robust exciton inding energy in aggregated structure-sorted carbon nanotubes revealed by two-photon excitation spectroscopy, {ACS Nano} \textbf{19}, 41252 (2025).
	
	\bibitem{Ando}T. Ando, Theory of electronic states and transport in carbon nanotubes, {J. Phys. Soc. Jpn.} \textbf{74}, 777 (2005).
	
	\bibitem{Bondarev12}I. V. Bondarev, Single-wall carbon nanotubes as coherent plasmon generators, {Phys. Rev. B} \textbf{85}, 035448 (2012).
	
	\bibitem{Bondarev12pss}I. V. Bondarev and T. Antonijevic, Surface plasmon amplification under controlled exciton-plasmon coupling in individual carbon nanotubes, {Phys. Stat. Sol. C} \textbf{9}, 1259 (2012).
	
	\bibitem{Bondarev14}I. V. Bondarev and A. V. Meliksetyan, Possibility for exciton Bose-Einstein condensation in carbon nanotubes, {Phys. Rev. B} \textbf{89}, 045414 (2014).
	
	\bibitem{Bondarev09}I. V. Bondarev, L. M. Woods, and K. Tatur, Strong exciton-plasmon coupling in semiconducting carbon nanotubes, {Phys. Rev. B} \textbf{80}, 085407 (2009).
	
	\bibitem{Pichler98}T. Pichler, M. Knupfer, M. S. Golden, J. Fink, A. Rinzler, and R. E. Smalley, Localized and delocalized electronic states in single-wall carbon nanotubes, {Phys. Rev. Lett.} \textbf{80}, 4729 (1998).
	
	\bibitem{WelschQO}W. Vogel and D.-G. Welsch, \emph{Quantum Optics} (Wiley-VCH, 2006). Ch. 10, p.337.
	
	\bibitem{BondLamb06}I. V. Bondarev and P. Lambin, Near-field electrodynamics of atomically doped carbon nanotubes, in: \emph{Trends in Nanotubes Research}, ed. D.A.Martin (Nova Science, NY, 2006), Ch.6, p.139.
	
	\bibitem{BondLamb05}I. V. Bondarev and P. Lambin, van der Waals coupling in atomically doped carbon nanotubes, {Phys. Rev. B} \textbf{72}, 035451 (2005).
	
	\bibitem{BondLamb04}I. V. Bondarev and Ph. Lambin, Spontaneous-decay dynamics in atomically doped carbon nanotubes, {Phys. Rev. B} \textbf{70}, 035407 (2004).
	
	\bibitem{Cardona}P. Y. Yu and M. Cardona, \emph{Fundamentals of Semiconductors}, 4th edn. (Springer-Verlag, 2010).
	
	\bibitem{Berest}V. B. Berestetskii, E. M. Lifshitz, and L. P. Pitaevskii, \emph{Qua\-n\-tum Electrodynamics} (Pergamon, 1982).
	
	\bibitem{AbersQM}E. S. Abers, \emph{Quantum Mechanics} (Pearson, NJ, 2004).
	
	\bibitem{BliokhAiello13}K. Y. Bliokh and A. Aiello, Goos-H\"{a}nchen and Imbert-Fedorov beam shifts: an overview, {J. Opt.} \textbf{15}, 014001 (2013).
\end{thebibliography}
\end{document}